\documentclass[aps,physrev,superscriptaddress,notitlepage,twocolumn]{revtex4-2}

\usepackage{graphicx}
\usepackage{amsmath,amssymb,amsfonts}
\usepackage{braket}
\usepackage{comment}
\usepackage{xcolor}
\usepackage{url}
\usepackage[hidelinks]{hyperref}
\usepackage{natbib}
\usepackage{dcolumn}
\usepackage{bm}
\usepackage{lineno}

\newcommand{\JQI}{Joint Quantum Institute and Joint Center for Quantum Information and Computer Science, University of Maryland and NIST, College Park, MD 20742 USA}

\begin{document}


\title{Observation of Stark many-body localization without disorder}

\author{W. Morong}
\email{wmorong@umd.edu}
\affiliation{\JQI}
\author{F. Liu}
\email{fliu1235@umd.edu}
\affiliation{\JQI}
\author{P. Becker}
\affiliation{\JQI}
\author{K. S. Collins}
\affiliation{\JQI}
\author{L. Feng}
\affiliation{\JQI}
\author{A. Kyprianidis}
\affiliation{\JQI}
\author{G. Pagano}
\affiliation{Department of Physics and Astronomy, Rice University, Houston, TX  77005 USA}
\author{T. You}
\affiliation{\JQI}
\author{A. V. Gorshkov}
\affiliation{\JQI}
\author{C. Monroe}
\affiliation{\JQI}

\begin{abstract}
Thermalization is a ubiquitous process of statistical physics, in which a physical system reaches an equilibrium state that is defined by a few global properties such as temperature.
Even in isolated quantum many-body systems, limited to reversible dynamics, thermalization typically prevails \cite{Rigol2008}. However, in these systems, there is another possibility: many-body localization (MBL) can result in preservation of a non-thermal state \cite{Abanin2019,Nandkishore2014}. While disorder has long been considered an essential ingredient for this phenomenon, recent theoretical work has suggested that a quantum many-body system with a spatially increasing field---but no disorder---can also exhibit MBL \cite{VanNieuwenburg2019}, resulting in `Stark MBL' \cite{Schulz2019}. Here we realize Stark MBL in a trapped-ion quantum simulator and demonstrate its key properties: halting of thermalization and slow propagation of correlations. Tailoring the interactions between ionic spins in an effective field gradient, we directly observe their microscopic equilibration for a variety of initial states, and we apply single-site control to measure correlations between separate regions of the spin chain. Further, by engineering a varying gradient, we create a disorder-free system with coexisting long-lived thermalized and nonthermal regions. The results demonstrate the unexpected generality of MBL, with implications about the fundamental requirements for thermalization and with potential uses in engineering long-lived non-equilibrium quantum matter.
    
\end{abstract}

\maketitle

\clearpage

Many-body localization was first formulated as a generalization of the Anderson transition \cite{Anderson1958,Lee1985,Gornyi2005,Basko2006}. In disorder, non-interacting quantum particles can experience destructive interference through multiple scattering, causing a transition to exponentially localized wavepackets. Over time, a cohesive picture of MBL in interacting systems has also developed \cite{Serbyn2013,Huse2014}. In this description, the MBL regime has extensive local conserved quantities that generalize the particle occupancies in Anderson localization. However, interactions result in additional slow spreading of correlations via entanglement. Strikingly, MBL creates a phase of matter that is non-ergodic: for a range of parameters, local features of the initial state are preserved for all times, preventing thermalization \cite{Abanin2019}.

In considering MBL, it is natural to ask whether random disorder is a requirement. A partial answer has long been known: MBL is possible with incommensurate periodic potentials \cite{Iyer2013}. However, the question of whether an MBL phase might exist which preserves translational symmetry, for instance in a system with gauge invariance \cite{Brenes2018} or multiple particle species \cite{Grover2014,Yao2016}, has continued to generate extensive discussion \cite{Alet2018}. Recently, this problem has been approached from a different starting point: the Bloch oscillations and Wannier-Stark localization of non-interacting particles in a uniformly tilted lattice \cite{Wannier1962}. From this, it has been predicted that interacting systems with a large linear tilt can also display MBL-like behavior \cite{VanNieuwenburg2019,Schulz2019}. This effect, sometimes called Stark MBL, has attracted considerable theoretical and experimental interest \cite{Taylor2019,Kshetrimayum2020,Zhang2020,Chanda2020,Bhakuni2020,Doggen2020,Khemani2020,Yao2021,Guardado-Sanchez2020,Scherg2020,Guo2020a}. However, clear experimental realization of Stark MBL has been complicated by approximate Hilbert space fragmentation that occurs in the limit of short-range interactions \cite{Scherg2020, Khemani2020, Sala2020}. The setting of a trapped-ion quantum simulator with long-range spin-spin couplings naturally overcomes this complication.

\section*{Experimental setup}

\begin{figure*}[!htb]
\centering
\includegraphics[width= 0.95 \textwidth]{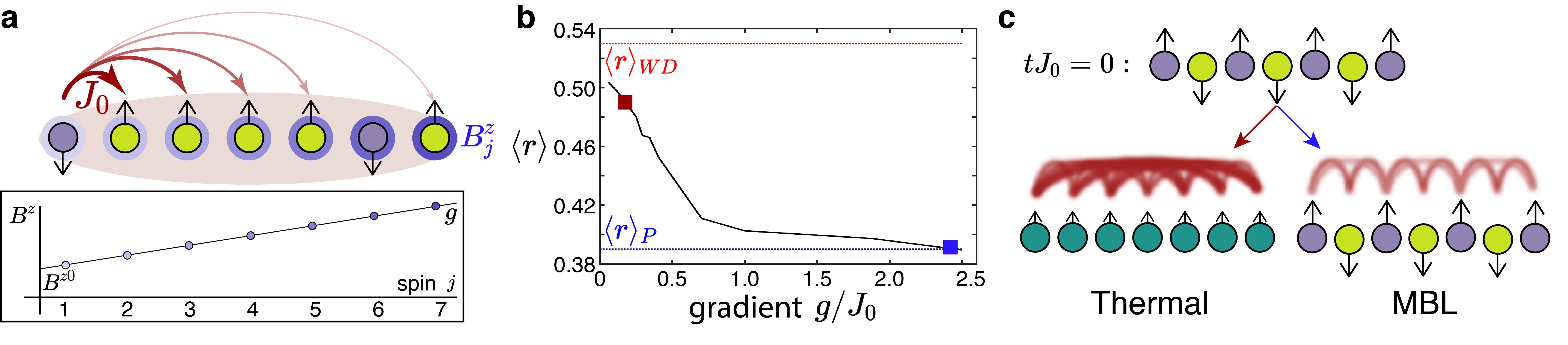}
\caption{Experimental setup. \textbf{a}, Each trapped ion in a chain of length $N$ encodes a pseudospin. Global lasers controllably mediate a long-range spin-spin coupling (red), which is parameterized by the nearest-neighbor rate $J_0$. A tightly focused beam provides a site-resolved effective $B^z$ magnetic field (blue) that is used to engineer a field gradient with slope $g$. For clarity, we show $N=7$. \textbf{b}, The parameter $\langle r \rangle$, a measure of the level statistics of the experimental Hamiltonian ($N=15$), shows a progression from statistics near the Wigner-Dyson limit ($\langle r\rangle_{WD}$, red dotted line) at small $g/J_0$, characteristic of a generic ergodic system, to Poisson statistics ($\langle r\rangle_{P}$, blue dotted line) at large $g/J_0$, characteristic of a localized system (see Extended Data Fig.~\ref{fig:FigHistogram} for full histograms of $r$). \textbf{c}, We probe the system using a quench from a non-equilibrium initial state, such as the N\'eel state shown here. At small $g/J_0$, an initial spin pattern will quickly relax to a uniform average magnetization, while at large $g/J_0$ the initial pattern persists. The former is consistent with a thermal state, in which uniformity is combined with entanglement (red links) reaching across the entire chain, while the latter is consistent with many-body localization, in which the magnetization remains non-uniform and entanglement spreads slowly.}
\label{fig:Fig1}
\end{figure*}

Investigation of many-body localization has been driven in part by the development of isolated quantum simulator platforms capable of single-site manipulation and readout \cite{Smith2016, Xu2018b, Chiaro2019, Brydges2019}. Our experimental apparatus (Fig.~\ref{fig:Fig1}a) consists of a chain ($N=$ 15 to 25) of $^{171}$Yb$^+$ ions, with pseudospin states $| \!\! \uparrow_z\rangle$ and $| \!\! \downarrow_z \rangle$ encoded in hyperfine levels. The Hamiltonian has two ingredients. The first is an overall spin-spin interaction, mediated by global laser beams coupling spin and motion using the M\o lmer-S\o rensen scheme \cite{Molmer1999}. The second is a programmable effective $B^z$ magnetic field at each ion, generated using a tightly focused beam \cite{Lee2016a}. Together these result in a versatile tool to study many-body physics. In addition to turning on or off either Hamiltonian term, we use the tightly focused beam to initialize spins in arbitrary product states, and we measure arbitrary local observables with state-dependent fluorescence collected onto a charge-coupled device (CCD) camera. 

Combining these terms and choosing the local field to be a linear gradient results in a tilted long-range Ising Hamiltonian ($\hbar=1$):
\begin{equation}
H=\sum_{j<j'} J_{jj'} \sigma_{j}^{x} \sigma_{j'}^{x}+\sum_{j=1}^N (B^{z0}+(j-1)g) \sigma_{j}^{z}.
\label{eq:ExpHam}
\end{equation}
Here we have the long-range spin-spin couplings $J_{jj'}$, approximately following a power-law: $J_{jj'} \approx J_0/|j-j'|^{\alpha}$, with $J_0$ the nearest-neighbor coupling and $\alpha=1.3$. $B^{z0}$ is an overall bias field, and $g$ the gradient strength, with \{$J_0$, $B^{z0}$, $g\}>0$. In practice, we generate this Hamiltonian stroboscopically, using a Trotterization scheme to reduce decoherence (see Methods and Extended Data Fig.~\ref{fig:FigTrotter}). The bias field $B^{z0}$ is set to be large ($B^{z0}/ J_0>5$), so that the total magnetization $\sum_j \langle \sigma^z_j \rangle$ is approximately conserved. With this constraint, and neglecting edge effects, $J_{jj'}=J_{|j-j'|}$ and the Hamiltonian is translationally invariant: the operation $j\rightarrow j+n$ for integer $n$ is equivalent to a shift in $B^{z0}$, which has no effect in the bulk.

With a disordered $B^z$ field, this system has been used to study MBL \cite{Smith2016}. For an initial state of definite total magnetization, the spin model can be mapped to a chain of hard-core bosons with long-range hopping in a potential (see Methods), indicating that it has similar ingredients to models previously used to study Stark MBL \cite{VanNieuwenburg2019, Schulz2019}.

A useful numeric diagnostic of whether a model exhibits an MBL regime can be found in the level statistics, which feature similar behavior in regular (disordered) MBL \cite{Oganesyan2007} and Stark MBL \cite{VanNieuwenburg2019,Schulz2019}. The energy levels of a generic thermalizing ergodic system follow the Wigner-Dyson distribution characterizing random matrices, while a generic many-body localized system has a Poissonian level distribution \cite{Oganesyan2007}. This difference can be quantified by the average ratio of adjacent energy level gaps, defined as
\begin{equation}
\langle r \rangle=\frac{1}{n}\sum_n\frac{\text{min}(E_{n+1}-E_n,E_n-E_{n-1})}{\text{max}(E_{n+1}-E_n,E_n-E_{n-1})}.
\end{equation}
The quantity $\langle r\rangle$ is 0.53 for a Wigner-Dyson distribution and 0.39 in the Poissonian case. Diagonalizing the Hamiltonian (Eq.~\ref{eq:ExpHam}) for $N=15$, we find that $\langle r \rangle$ moves from 0.50 to 0.39 as the gradient $g/J_0$ is increased, suggesting increasing localization (Fig.~\ref{fig:Fig1}b). While Fig.~\ref{fig:Fig1}b shows the exact experimental Hamiltonian, including deviations from uniform couplings near the edges of the chain, this behavior persists in a uniform Hamiltonian (see Methods and Extended Data Figs.~\ref{fig:FigPD},\ref{fig:FigScaling}). Unlike previous studies of Stark MBL, in which a small amount of disorder or curvature was required for Poissonian level statistics \cite{VanNieuwenburg2019,Schulz2019}, Eq.~\ref{eq:ExpHam} exhibits them without any terms perturbing the translational symmetry.

We probe the degree of localization using a quench procedure, shown schematically in Fig.~\ref{fig:Fig1}c. The initial state, such as a N\'eel state of staggered up and down spins, is typically highly excited and far-from-equilibrium. If it thermalizes, the dynamics following the quench will lead to a state in which each spin has a uniform probability of being up or down. Many-body localization will instead result in persisting memory of the initial configuration, breaking ergodicity.

\section*{Non-thermalization from Stark MBL}

\begin{figure*}[!htb]
\centering
\includegraphics[width= 0.8 \textwidth]{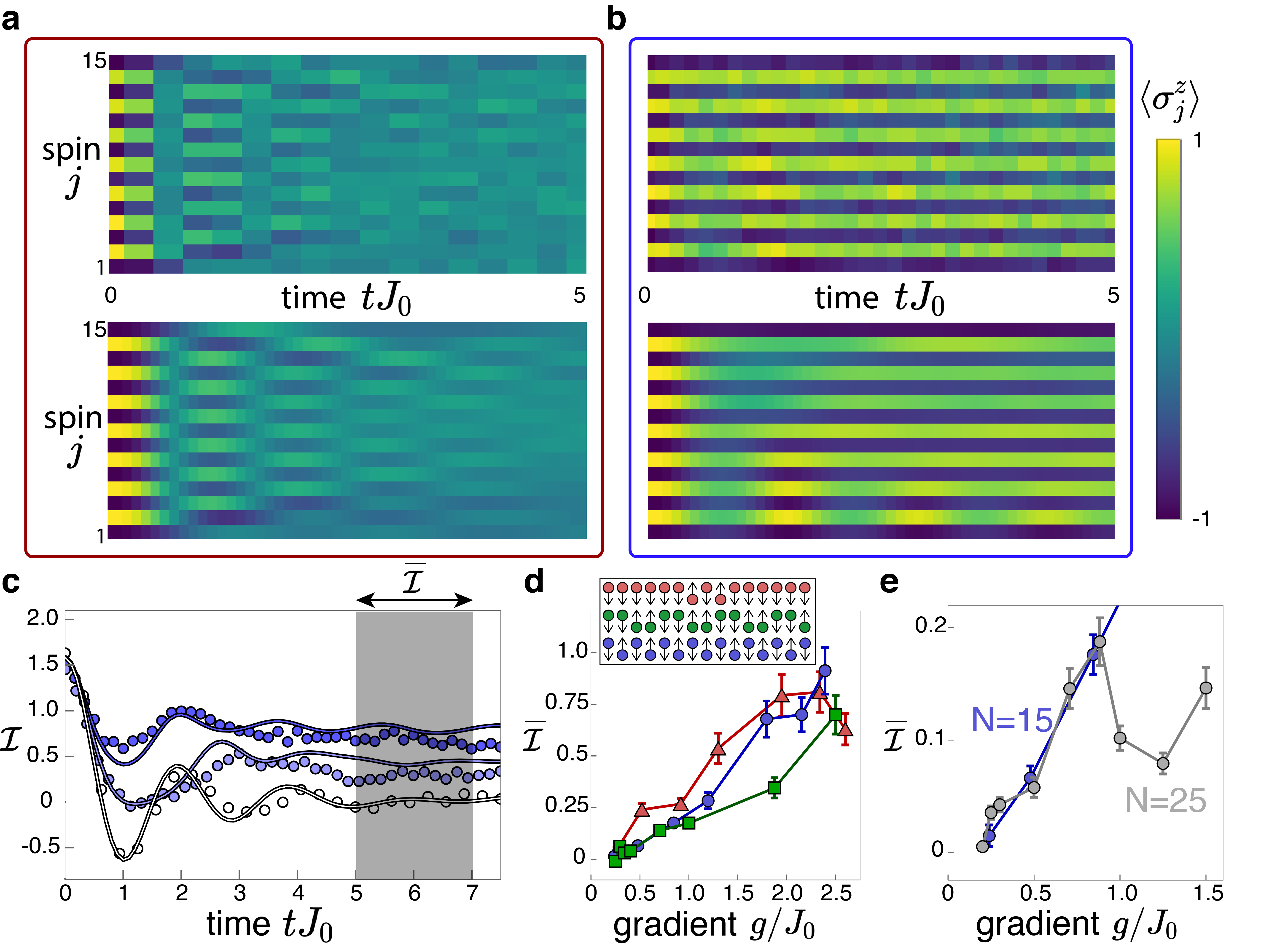}
\caption{Non-thermalization from Stark MBL. \textbf{a}, Ion-resolved dynamics for an initial N\'eel state ($N=15$) at $g/J_0=0.24$, and \textbf{b}, at $g/J_0=2.4$, corresponding to the red and blue points on Fig.~\ref{fig:Fig1}b. While the state quickly relaxes to a uniform magnetization in the small gradient, the large gradient results in a persisting memory of the initial state. The top row is experimental data, averaged over 200 repetitions, and the bottom row is exact numerics. \textbf{c}, Memory of the initial state, here a N\'eel state ($N=15$), can be quantified by the generalized imbalance $\mathcal{I}$. For a state of frozen up and down spins, $\mathcal{I}=2$, and for complete relaxation to a uniform state, $\mathcal{I}=0$. As the gradient is increased (light to dark), the imbalance crosses from quick relaxation towards zero to a persistent finite value. Points are experimental data at $g/J_0=$ \{0.24, 1.2, 1.8\}, with statistical error bars smaller than the symbol size, and lines are exact numerics using the experimental Hamiltonian. Numerics in \textbf{a}, \textbf{b}, and \textbf{c} incorporate experimental noise (see Methods and Extended Data Fig.~\ref{fig:FigNoise}). \textbf{d}, For various initial states, shown at top, we see a similar value of the late-time imbalance at large gradient, suggesting uniform localization. From top to bottom, the three initial states correspond to the \{triangle, square, round\} points. \textbf{e}, Dependence of the late-time imbalance on system size is shown, using an initial N\'eel state with $N=15$ (a subset of the data in panel d) and $N=25$. The overall increase of late-time imbalance with gradient is robust to the system size increase. The pronounced dip in $\overline{\mathcal{I}}$ near $g/J_0=1.0$ may be partly due to a finite-time feature that appears near this value (see Methods and Extended Data Fig.~\ref{fig:FigFinizeSizeImbal}). Error bars throughout represent statistical uncertainty of the mean value (1$\sigma$ s.e.m.).}
\label{fig:Fig2}
\end{figure*}

Performing the quench experiment, we see the expected signature of localization: a small gradient results in quick equilibration of the spins (Fig.~\ref{fig:Fig2}a), while in a large gradient they remain near their initial values over the experimental timeframe (Fig.~\ref{fig:Fig2}b). The experimental data correspond closely to exact numerical simulations.

To quantify the amount of initial state memory, we define a generalized imbalance, $\mathcal{I}(t)$. This observable is similar to other previously used measures of initial state memory, such as the imbalance \cite{Schreiber2015} or the Hamming distance \cite{Smith2016}, but is advantageous for comparing different initial states (see Methods). For an initial state with $M$ spins that are up, and $N-M$ down, $\mathcal{I}$ is equal to the subsequent difference between the average magnetizations of the two groups:
\begin{equation}
\mathcal{I}(t)=\frac{\sum_{j=1}^M\langle \sigma_j^z(t)\rangle}{M}-\frac{\sum_{j'=1}^{N-M}\langle \sigma_{j'}^z(t)\rangle}{N-M} 
\end{equation}
where $j$ ($j'$) only sums over the spins that were initialized up (down). In general, $|\mathcal{I}(t)|$ varies from two, for perfect memory of an initial state composed of up and down spins, to zero, for a uniform state as at thermal equilibrium.

The imbalance shows a clear trend as we increase the gradient (Fig.~\ref{fig:Fig2}c). At smaller gradients, it relaxes to a decaying oscillation about zero, indicating quick thermalization. However, as the gradient is increased, the imbalance instead settles to a progressively higher value. Compared to exact numerics, decoherence causes a slow decay of $\mathcal{I}$ over time, attributed primarily to residual excitation of ion-chain motion. However, the separation between this decoherence time and the fast relaxation dynamics allows us to characterize the late-time imbalance.

To study initial-state memory for different gradients, we average $\mathcal{I}(t)$ over a time window $tJ_0$ from 5 to 7. This window is chosen to be late enough that transient oscillations have largely decayed, while early enough that decoherence is limited. This late-time imbalance, $\bar{\mathcal{I}}$, captures the amount of initial-state memory remaining after any initial relaxation, and thus the approximate degree of localization (Fig.~\ref{fig:Fig2}d). At the smallest gradient $\bar{\mathcal{I}}$ is consistent with zero: averaging over the initial states shown in Fig.~\ref{fig:Fig2}d we have $\bar{\mathcal{I}}=0.017\pm0.027$ (1$\sigma$ s.e.m.). With a larger gradient, $\bar{\mathcal{I}}$ becomes clearly distinct from zero and progressively increases, reflecting increasing memory of the initial state. Crucially, this memory does not show strong dependence on the specific initial state: for states with different numbers of initial spin flips and different symmetry properties, similar behavior is observed. This initial state insensitivity is consistent with many-body localization, which can have some energy dependence \cite{Zhang2020} but is a robust mechanism for breaking ergodicity that can span the entire spectrum. This can be contrasted with other effects that do cause thermalization to be strongly state-dependent, such as quantum many-body scars \cite{Bernien2017} and domain wall confinement \cite{Tan2019}. However, compared to disordered MBL, a key difference is also evident: a small nonzero value of $\bar{\mathcal{I}}$, and thus imperfect thermalization, persists at small values of $g/J_0$. This is consistent with the expectation that even in this regime thermalization is anomalously slow or incomplete \cite{Gromov2020, Doggen2020}, which progresses towards complete localization as the gradient increases.

A key further test of Stark MBL is to characterize its dependence on increasing system size. This is especially relevant to systems with long-range terms, where finite-size effects may be particularly important \cite{Smith2016,Wu2016}. Increasing the length to $N=25$, we see a rise in the imbalance at small $g/J_0$ that is similar to the $N=15$ case (Fig.~\ref{fig:Fig2}e).
While we are unable to reach the deeply localized regime for $N=25$ due to the scaling of the experimentally achievable maximum gradient with $N$ (see Methods), the small nonzero value of $\overline{\mathcal{I}}$ that we observe indicates the persistence of a Stark MBL regime.

\section*{Revealing the correlated Stark MBL state}
 
\begin{figure}[!htb]
\centering
\includegraphics[width=0.5 \textwidth]{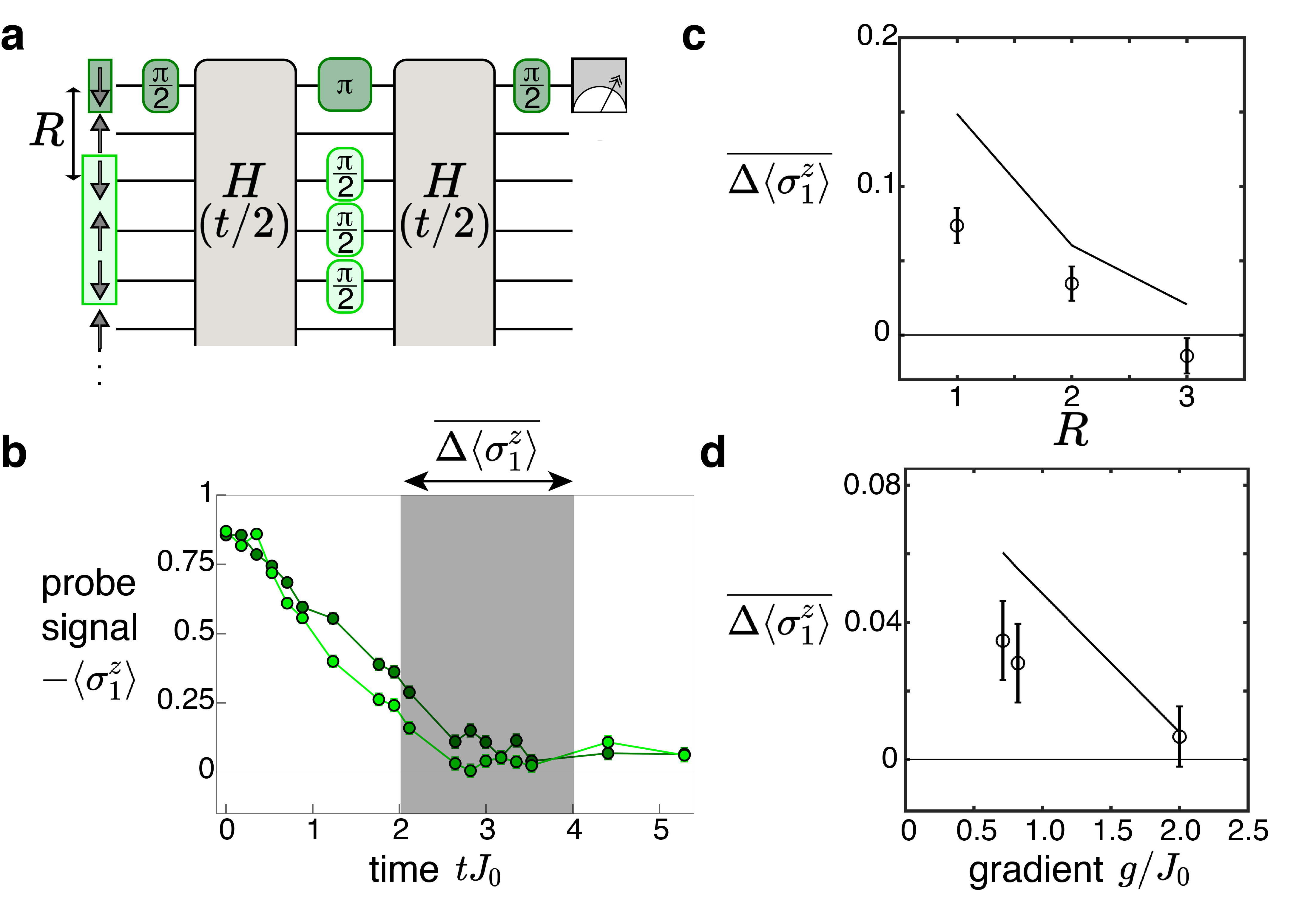}
\caption{DEER Protocol. \textbf{a}, In the spin-echo procedure (dark green), a single probe spin undergoes a spin-echo sequence, while the rest of the spins experience normal evolution under $H$ for total time $t$. In the DEER procedure (dark and light green) there are additional perturbing $\pi/2$ pulses on a region, here fixed at a size of three spins, that is $R$ spins away (with the case of $R=2$ shown). The difference in the probe magnetization following these procedures reflects the ability of the DEER region to influence the dynamics at the probe spin. We study this protocol using an initial N\'eel state ($N=15$). \textbf{b}, At intermediate times, before the spin-echo signal approaches zero due to decoherence, a difference develops between the spin-echo (dark green) and DEER (light green) signals. We quantify this by taking the average difference (DEER-spin echo) between $tJ_0=2$ and 4 (shaded region) after imbalance dynamics have stabilized. These data are for $R=1$ and $g/J_0=0.71$, and are averaged over 2000 repetitions.  \textbf{c}, As $R$ is increased (at $g/J_0=0.71$), the difference signal drops to zero, reflecting the incomplete spread of correlations through the system at finite time. \textbf{d}, As $g$ is increased (at $R=2$), the difference signal also decreases with increasing gradient, consistent with the expectation that within the Stark MBL phase, increasing localization leads to progressively slower development of correlations. Points in \textbf{c} and \textbf{d} are the experimental data, and solid lines are exact numerics incorporating experimental noise (see Methods and Extended Data Fig.~\ref{fig:FigNoise}).}
\label{fig:Fig3}
\end{figure}

Probes of the local magnetization, as in Fig.~\ref{fig:Fig2}, can establish non-thermalization over experimental timeframes, but they do not reveal the correlations that characterize a localized phase. The structure of the regular MBL phase is understood to be defined by emergent local conserved quantities \cite{Serbyn2013,Huse2014}. However, the resulting localized regions still interact with one another, leading to spreading of correlations via entanglement after a quench from a product state (typically logarithmic spreading in time, but potentially faster for long-range systems \cite{Pino2014, Safavi-Naini2019}). While the existence of these conserved quantities in Stark MBL is debated \cite{Doggen2020,Khemani2020}, there are indications that it can display similar entanglement dynamics \cite{Schulz2019,Taylor2019}.

Some observables have been established to directly probe this correlation spreading, such as quantum Fisher information \cite{Smith2016,Guo2020a} (see Methods and Extended Data Fig.~\ref{fig:FigQFI}) or techniques to measure subsystem entanglement entropy \cite{Lukin2019,Brydges2019}. We instead adopt a local interferometric scheme, the double electron-electron resonance (DEER) protocol, to reveal the spread of correlations \cite{Serbyn2014,Taylor2019,Chiaro2019}. This protocol, shown in Fig.~\ref{fig:Fig3}a, compares two experimental sequences: one that is a standard spin-echo sequence on a probe spin within a system of interest, and one that combines this with a set of $\pi/2$-pulse perturbations on a separate subregion, the `DEER region'. The spin-echo sequence cancels out static influences on the probe spin, either from global external fields or from fixed configurations of the surrounding spins. If this cancellation is perfect, the probe spin will return to its initial magnetization. The DEER sequence, by contrast, removes this cancellation for any effect of the spins in the DEER region on the probe. As a result, a difference in the final probe magnetization between the two sequences reflects correlations between the probe and DEER region generated by the dynamics. At sufficiently long times, a difference between these signals will develop in an MBL phase, but not in a thermal or non-interacting localized phase. This differential measurement setup is also naturally robust against common-mode non-idealities, including experimental noise. As this protocol requires control of the Hamiltonian and single-site manipulation and readout, it demonstrates how the capabilities of our experimental platform can enable methods of characterizing many-body systems beyond typical observables.

In Fig.~\ref{fig:Fig3}b-d, we demonstrate the DEER protocol and apply it to characterize the Stark MBL regime. Over time, a difference accumulates in the probe magnetization following the two procedures, reflecting the spread of correlations (Fig.~\ref{fig:Fig3}b). These correlations continue to move through the system after the imbalance has stabilized (see Methods and Extended Data Fig.~\ref{fig:FigDEERCompare}), indicating that they are not solely due to transient dynamics. Picking a time range after this initial evolution, $tJ_0=$2 to 4, we characterize the correlations by taking the average difference between the signals, $\overline{\Delta \langle \sigma^z_1 \rangle}$. This time window is slightly earlier than the window used for the steady-state imbalance, as the DEER signal is more sensitive to fluctuations in the local effective $B^z$ fields, which are the dominant source of experimental noise (see Methods). Varying the DEER spin distance, $R$, we see that this difference signal decreases for a DEER region farther from the probe, reflecting the local nature of correlation propagation (Fig.~\ref{fig:Fig3}c). Similarly, at a fixed separation and time window, we observe the reduction of the difference signal with increasing gradient, confirming that the correlation spread is controlled by the degree of localization (Fig.~\ref{fig:Fig3}d). The dependence of the difference signal on both $R$ and $g/J_0$ track exact numerics, with an overall scaling difference due to decoherence reducing the experimental signal. Taken together, these probes identify the Stark MBL regime as one in which correlations spread slowly through the system despite persisting memory of the initial state, distinguishing it from non-interacting localization.

\section*{Disorder-free MBL beyond a linear field}

\begin{figure*}[!htb]
\centering
\includegraphics[width= 0.9 \textwidth]{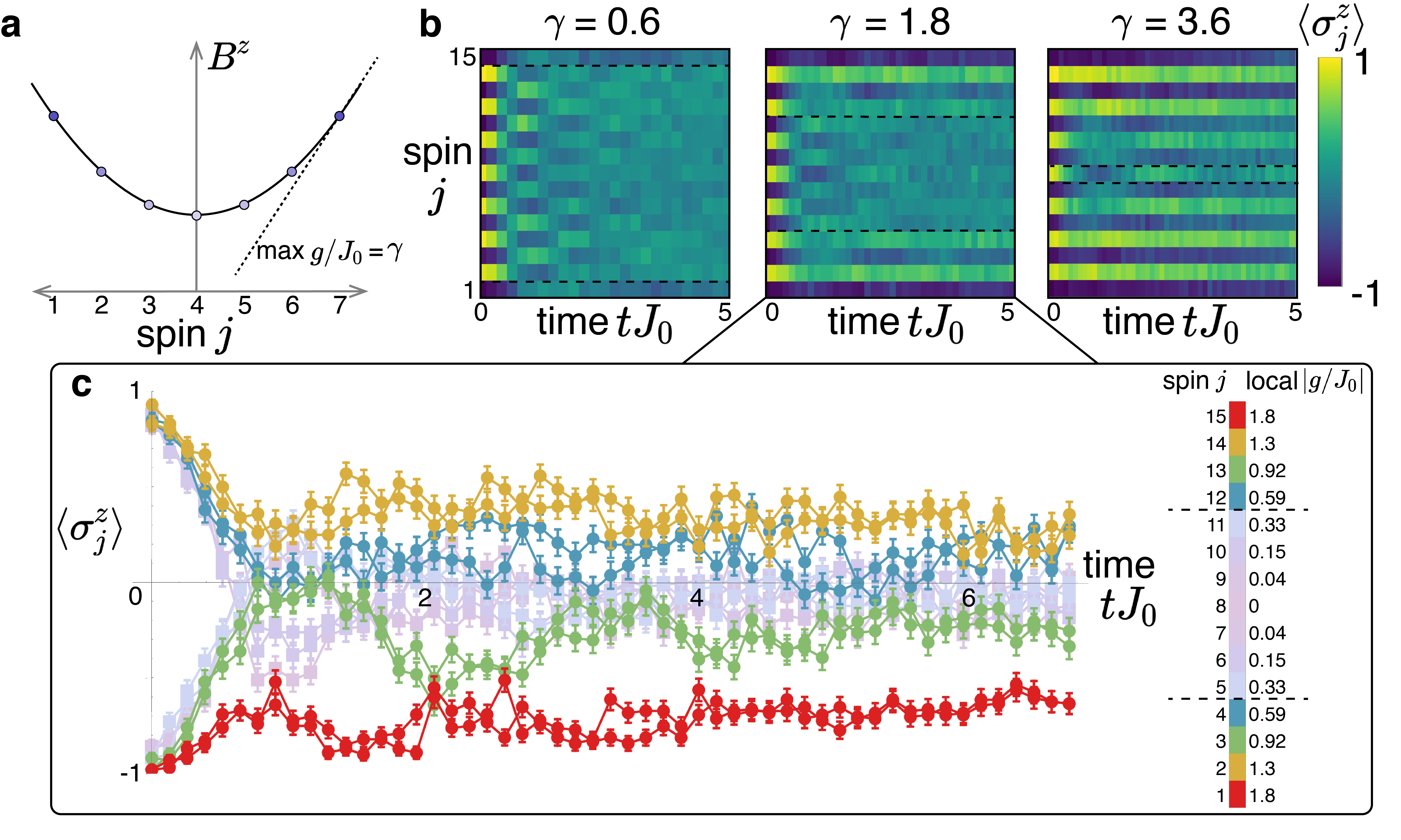}
\caption{Relaxation in a quadratic field. \textbf{a}, We reconfigure the site-resolved field from a linear gradient to a quadratic, characterized by the maximum slope $\gamma$.  For clarity, we show $N=7$. \textbf{b}, Dynamics are split into a thermalizing region near the center of the system and localized regions near the edges, with the approximate boundaries indicated by the dashed lines. As the maximum gradient is increased, the fraction of the system in the thermalizing regime shrinks. \textbf{c}, Ion-resolved traces of the dynamics for $\gamma=1.8$, showing separation of the spins into localizing regions (bright hues with round points) and thermalizing regions (faded hues with square points). Colors reflect the local field strength at each ion. Data are averaged over 200 repetitions.}
\label{fig:Fig4}
\end{figure*}

If many-body localized effects appear in a linearly increasing field, might they also be seen in a more general class of smoothly varying fields? Utilizing the tunability of this simulator, we investigate a natural generalization: a quadratic, rather than linear, potential. We parameterize the Hamiltonian as:
\begin{equation}
H=\sum_{j<j'} J_{jj'} \sigma_{j}^{x} \sigma_{j'}^{x}+\sum_{j=1}^N \left(B^{z0}+\frac{\gamma J_0 (j-\frac{N+1}{2})^2}{N-1}\right) \sigma_{j}^{z}.
\label{eq:Quadratic}
\end{equation}
Eq.~\ref{eq:Quadratic} describes a quadratic effective $B^z$ field with a minimum in the center of the system and a maximum slope of $g/J_0 = \pm \gamma$ at the ends of the chain. Similar models have been predicted to feature a spatial separation into an ergodic core and many-body localized edges \cite{Chanda2020}.

We summarize the results in Fig.~\ref{fig:Fig4}. Taking an initial N\'eel state ($N=15$), we observe a separation of the spins into thermalizing and localized regions, which appear to evolve independently. We determine an approximate dividing line between these regions as the innermost spins that are clearly distinct from the thermalizing region. For a range of curvatures $\gamma<3.6$, this occurs at a local slope of $g/J_0\approx 0.5$ (see Methods and Extended Data Fig.~\ref{fig:FigQuadCriticalGradient}), comparable to observations in Fig.~\ref{fig:Fig2}.

The comparison between Stark MBL in a linear gradient and disorder-free MBL in a quadratic field suggests similar localizing mechanisms. While a large gradient results in a model with approximate center-of-mass (or dipole) conservation, a quadratic field instead results in a quadrupole constraint. Fractonic models in these limits display dynamics determined by the type of conservation, such as characteristic subdiffusion \cite{Gromov2020,Guardado-Sanchez2020}. However, our realizations of disorder-free MBL are far from these limits of exactly conserved moments, and over experimentally relevant times appear to exhibit dynamics that are determined by the local potential landscape, rather than overall constraints \cite{Chanda2020, Yao2021}.

The quadratic field is also an intriguing venue to explore the stability of disorder-free MBL in proximity to a thermalizing region. In regular MBL, it is believed that a thermal inclusion can induce many-body avalanches that slowly destabilize the MBL region \cite{DeRoeck2017,Leonard2020}. Disorder-free MBL, which does not feature any resonances between sites, may evade this instability. The observation of a localized region in a quadratic field is also directly relevant to longstanding questions about the state of correlated ultracold atoms in an optical lattice with harmonic confinement \cite{Kondov2013a}.

\section*{Discussion}

We have seen the signatures of many-body localization in a system without disorder, suggesting that the concept of MBL may be relevant in settings beyond the original considerations \cite{Gornyi2005,Basko2006}. For all types of MBL, questions remain about the conditions for asymptotic stability, particularly in systems with long-range terms or more than one dimension \cite{DeRoeck2017,Agarwal2017,VanNieuwenburg2019}. To this end, future work could study the dependence of Stark MBL on the coupling range $\alpha$, or explore connections between our observations and the approximate Hilbert space fragmentation (or shattering) arising in certain short-range tilted models \cite{Sala2020,Khemani2020,Doggen2020,Taylor2019, Moudgalya2019} (see Methods).

Extension of Stark MBL to the thermodynamic limit poses several challenges. Infinite energy differences appear between different ends of the system, although this can be partially addressed through a gauge transformation recasting the gradient as a time-dependent drive (see Methods). 
Furthermore, slow state-dependent processes can result in increasing delocalization with system size (see Methods and Extended Data \ref{fig:ExactDiag}).
However, localization can be extended to arbitrarily long times by increasing the field gradient, adding gradient curvature, or restriction to finite sizes.

Finally, from the perspective of near-term quantum devices our results suggest that Stark MBL retains key aspects of the disordered MBL phase while offering certain advantages, such as not requiring a fine-grained field or disorder averaging of observables.
Stark MBL may be a useful resource for such devices, serving as a tool to stabilize driven non-equilibrium phases \cite{Else2019,Kshetrimayum2020}, or as a means of making a quantum memory \cite{Nandkishore2014} with each site spectroscopically resolved.

\section*{Acknowledgements}
We acknowledge helpful discussions with Alan Migdall and Rahul Nandkishore.

This work is supported by the DARPA Driven and Non-equilibrium Quantum Systems (DRINQS) Program (D18AC00033), the NSF Practical Fully-Connected Quantum Computer Program (PHY-1818914), the DOE Basic Energy Sciences: Materials and Chemical Sciences for Quantum Information Science program (DE-SC0019449), the DOE High Energy Physics: Quantum Information Science Enabled Discovery Program (DE-0001893), the DoE Quantum Systems Accelerator, the DOE ASCR Quantum Testbed Pathfinder program (DE-SC0019040), the DoE ASCR Accelerated Research in Quantum Computing program (DE-SC0020312), the AFOSR MURI on Dissipation Engineering in Open Quantum Systems (FA9550-19-1-0399), and the Office of Naval Research (Award N00014-20-1-2695). The authors acknowledge the University of Maryland supercomputing resources made available for conducting the research reported in this work.

\section*{Data Availability}
The data that support the findings of this study are available from the corresponding author upon request.

\section*{Code availability}
The code used for analyses is available from the corresponding author upon request.

\bibliography{library.bib}

\section*{Author contributions}
F.L., L.F., and W.M. proposed the experiment. W.M., P.B., K.S.C., A.K., G.P., T.Y., and C.M. contributed to experimental design, data collection, and analysis. F.L. and A.V.G. contributed supporting theory and numerics. All authors contributed to the manuscript.

\section*{Competing interests}

The authors declare competing financial interests: C.M. is Co-Founder and Chief Scientist at IonQ, Inc.


\section*{Materials and correspondence}

Correspondence and requests for materials should be addressed to William Morong (\href{mailto:wmorong@umd.edu}{wmorong@umd.edu}) or Fangli Liu (\href{mailto:fliu1235@umd.edu}{fliu1235@umd.edu}).

\section*{Methods}










\renewcommand{\figurename}{Extended Data Figure}
\setcounter{figure}{0}

\section*{Experimental Apparatus}

\subsection*{State preparation and readout}

Our apparatus has been previously described in \cite{Islam2011,Zhang2017a,Zhang2017,Monroe2019a}. We employ a three-layer Paul trap to confine $^{171}$Yb$^+$ ions in a harmonic pseudopotential with trapping frequencies $f_{x,y}=$ 4.64 MHz and either $f_{z}=$ 0.51 MHz ($N=15$) or 0.35 MHz ($N=25$). There is a 1 \% to 2 \% day-to-day variation in these frequencies. Pseudospins are encoded in the two clock ground hyperfine states, with $|F=0,m_F=0\rangle=|\!\!\downarrow_z\rangle$ and $|F=1,m_F=0\rangle=|\!\!\uparrow_z\rangle$. We drive coherent global rotations between these spin states using stimulated Raman transitions. Long-range spin-spin interactions are generated via a bichromatic beatnote that couples these states via motional modes along the $\hat{x}$ direction. This is generated by three Raman beams from a pulsed 355 nm laser driving a symmetric pair of transitions, with average detunings of $\mu/2\pi=$ 200 kHz from the red and blue sideband transitions of the highest frequency (center-of-mass) transverse motional mode along $\hat{x}$. The resulting distribution of $J_{jj'}$ couplings has a best-fit power law of $\alpha=1.28$ for $N=15$ and $\alpha=1.31$ for $N=25$, and a best-fit $J_0/2\pi$ between 0.25 kHz and 0.33 kHz, depending on day-to-day variations in laser power. This value of $J_0$, calibrated for a given day, is used to scale energies and times in the main text.

Each experimental cycle begins with state initialization via optical pumping and Doppler and resolved-sideband cooling, which prepares the spin state $|\!\!\downarrow_z\rangle$ with fidelity $>0.99$ and the ground motional state with fidelity $>0.9$. Arbitrary product states are initialized using the site-dependent AC Stark shift from the individual addressing beam (from the same 355 nm light generating the Ising interactions), combined with overall rotations, with typical preparation fidelities of $>0.9$ per spin. Readout is performed via state-dependent fluorescence using the 369.5 nm $|\!\!\uparrow_z\rangle\rightarrow {}^2 P_{1/2}$ transition collected on a CCD camera, with typical detection errors of 3 \%. All measurements presented in the main text, except for the DEER measurements, are repeated at each setting 200 times for statistics. For the DEER measurements, we instead average over 2000 repetitions, which are taken alternating between DEER and spin-echo sequences every 100 measurements so that to a very good approximation both sample any noise profile equally. The data presented have not been corrected for state preparation and measurement (SPAM) errors.

\subsection*{Calibration of Hamiltonian parameters}

The experimental $J_{jj'}$ matrix is determined by measurements of motional sideband Rabi frequencies and trap parameters. Past work has validated this model against direct measurements of the matrix elements \cite{Smith2016}.

We directly measure and calibrate the linear field for each spin individually. As this calibration process is imperfect, each spin has a finite amount of deviation from the ideal linear gradient and thus there is a finite amount of effective site-by-site disorder in the experimental realization, with $\delta \frac{B^z_j}{gj} \approx 0.02$. While a small amount of disorder can be crucial in simulations of Stark MBL with short-range terms, because it breaks the exact degeneracies of that problem \cite{VanNieuwenburg2019}, in the context of long-range couplings the level statistics are already generic, and this disorder does not have a substantial effect on the system in numerics over experimental timeframes. As such, we call our system `disorder-free' in the sense that we only have small, technical and well-understood imperfections limiting our realization of the ideal disorder-free Hamiltonian. Any real quantum simulator can only hope to asymptotically approach a perfectly uniform environment, just as any quantum simulator can only hope to approximately realize MBL because there will always be some residual coupling to the environment that restores ergodicity at sufficiently long times.

\section*{Generalized Imbalance}

The generalized imbalance used in the main text is defined as:

\begin{align}
    \mathcal{I}(t)&=\frac{\sum_j \langle \sigma^z_j(t)\rangle (1+\langle \sigma^z_j(0)\rangle)}{\sum_j(1+\langle \sigma^z_j(0)\rangle)}\nonumber\\
    &-\frac{\sum_j \langle \sigma^z_j(t)\rangle (1-\langle \sigma^z_j(0)\rangle)}{\sum_j(1-\langle \sigma^z_j(0)\rangle)}
\end{align}

For an initial state that is a product of up and down spins along $z$, this reduces to a simple form: the average magnetization of the spins initialized up minus the average magnetization of the spins initialized down. For an initial state that is fully polarized this imbalance is undefined, which may be considered as a drawback to this measure, but such a state is already near equilibrium and thus is not useful for quantifying equilibration.

This definition is similar to many other variations of the imbalance. For an initial N\'eel state with an even number of spins it is identical up to scaling factors to both the imbalance and the Hamming distance, while for a general initial state of up and down spins it reduces to an alternate `generalized imbalance' that has been used in previous studies \cite{Wei2019,Guo2020,Guo2020a}. However, in general this definition offers a few advantages:
\begin{itemize}

\item Unlike the imbalance, it is exactly zero for a thermalized system with an odd number of spins.

\item It does not require any knowledge of the initial state to be added in by hand, unlike alternative observables in which the initially flipped spins are tracked.

\item Unlike the Hamming distance, this generalized imbalance is zero for a thermalized system, and has units of magnetization difference (therefore ranging from -2 to 2).

\item Finally, this generalized imbalance is less sensitive to some noise terms than the Hamming distance, such as spurious processes that do not conserve the overall magnetization. An example is useful: consider an initial state of one flipped spin ($\langle \sigma^z\rangle=1$), with $N=10$, and a background of spin-down ($\langle \sigma^z\rangle=-1$). Then, suppose that after some time this system has either evolved to a completely uniform system with an average magnetization of -1, or a state where each spin relaxes by 0.2 towards zero magnetization, leaving the initially flipped spin at a magnetization of +0.8 and the remaining spins at -0.8. Both of these final states have the same Hamming distance from the initial state of 0.1, because they both represent a system that is an average of one spin flip from the initial state. However, the first final state is completely equilibrated, while the second has a strong memory of the initial state. The Hamming distance, therefore, is not an optimal measure of initial state memory in a situation where a few flipped spins give you more information about the initial state than the background spins.

\end{itemize}

While the Hamming distance is always zero at time zero, this generalized imbalance only starts at 2 for an initial state in which each spin is in a definite state of $\sigma^z$. In Fig.~\ref{fig:Fig2}c the experimental imbalances do not start exactly at 2, reflecting SPAM errors.

\section*{Numerics}
Studies of Hamiltonian level statistics with $\langle r \rangle$ use exact diagonalization of the Hamiltonian. For simulations of dynamics when the chain length does not exceed $L=23$ we solve the Schr{\"o}dinger equation using the  Krylov space technique \cite{Luitz2017,Nauts1983}. For simulation of dynamics with $L=25$, we use the fourth-order Suzuki-Trotter expansion to decompose the Hamiltonian into two pieces, and use a global Hadamard transformation to rotate the basis of operators \cite{PhysRevLett.123.115701}. This reduces the memory required in the simulation since the Hamiltonian is diagonal (with the rotation) and does not need to be stored as a matrix form.

For all numerics, except those shown in the subsequent Methods section `Numerical studies of the ideal power-law Hamiltonian,' we use the experimentally determined $J_{jj'}$ matrix. These couplings show some inhomogeneity across the chain, with the nearest-neighbor hopping varying 7 \% for $N=15$. At large ion-ion separation they also show deviations from power-law behavior, with the couplings falling off faster than the best-fit power law \cite{Monroe2019a}. The comparison to power-law numerics shows that each of these effects does not strongly alter the dynamics.

\subsection*{Experimental noise model}

\begin{figure}[!htb]
\centering
\includegraphics[width= 0.45 \textwidth]{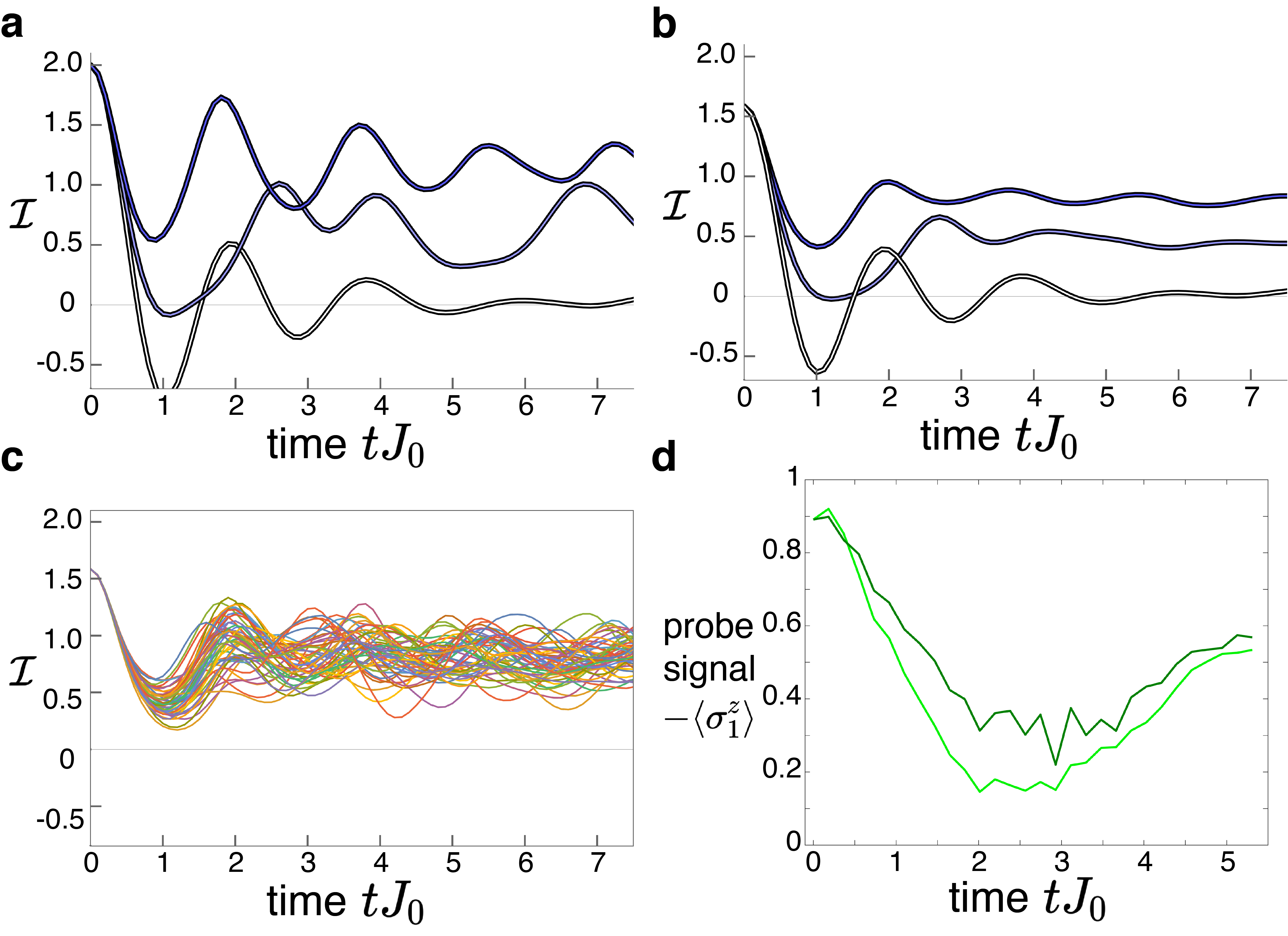}
\caption{Impact on noise model on dynamics. \textbf{a,b}, Noiseless (\textbf{a}) and noisy (\textbf{b}) numerics for an initial N\`eel state with $g/J_0=$ \{0.24, 1.2, 1.8\} (light to dark), corresponding to the data in Fig.~\ref{fig:Fig2}c. Compared to the ideal numerics, the noisy numerics show overall lower imbalances, primarily due to the SPAM errors, and damped oscillations, primarily due to variations in the individual local effective $B^z$ fields. However, these noise sources do not strongly affect the stability of the imbalance. \textbf{c}, Individual noisy realizations corresponding to the highest gradient shown above. \textbf{d}, Noise-averaged DEER simulations corresponding to Fig.~\ref{fig:Fig3}b.}
\label{fig:FigNoise}
\end{figure}

For numerics that are compared directly to experimental data in Fig.~\ref{fig:Fig2} and \ref{fig:Fig3} of the main text, we take the effects of experimental noise into account. We incorporate noise of the following types:
\begin{itemize}
    \item An error in the initial state, roughly accounting for the combined SPAM errors, consisting of a uniform rotation of the N\'eel state by $0.075 \pi$ radians in the Z-X plane.
    \item A shot-to-shot random variation of the overall field offset $B^{z0}$, with Gaussian variance of $2\pi \cdot 0.6$ kHz
    \item A shot-to-shot random variation of the gradient slope $g$, with a standard deviation of 6.25 \%
    \item A shot-to-shot random variation of the individual local field terms, deviating from the ideal linear gradient, with a standard deviation of 3.125 \% 
    
\end{itemize}

Each numerics line in Figs.~2 and 3 show the result of averaging over 50 random instances, drawn from Gaussian distributions of each parameter. In general, these error sources and magnitudes are consistent with independent estimations of our SPAM errors and laser intensity fluctuations. Notably, as the gradient is generated from a fourth-order Stark shift, the associated fractional noise is double that of the laser intensity fluctuations \cite{Lee2016a}. However, the precise values of the four error terms are chosen to match experiment. Owing to the large amount of data available, and the subtle differences in the effects of each term, these terms can be optimized fairly independently. For example, the noise in $g$ and the noise in the variation about $g$ for individual spins each give slightly different effects in the damping of the imbalance and the degree of asymmetry between small and large gradient. 

Extended Data Fig.~\ref{fig:FigNoise} shows a side-by-side comparison of the noiseless and noisy numerics for the imbalances shown in Fig.~\ref{fig:Fig2}, examples of the individual realizations that are averaged, and an example of the DEER signal.

Two error sources that are not included in the model are coupling to phonons and fluctuations of the local $B^z$ fields that occur during a single experimental run rather than from shot to shot. These are believed to dominate the remaining differences between experiment and theory, such as the slow decay of the experimental imbalance and the decay of the experimental DEER signal after $tJ_0 \approx 2$. However, the broad agreement observed in Fig.~\ref{fig:Fig2} and \ref{fig:Fig3} indicates that we have captured the main noise effects.

\section*{Trotterized M-S Hamiltonian}

\begin{figure*}[!htb]
\centering
\includegraphics[width= 0.9 \textwidth]{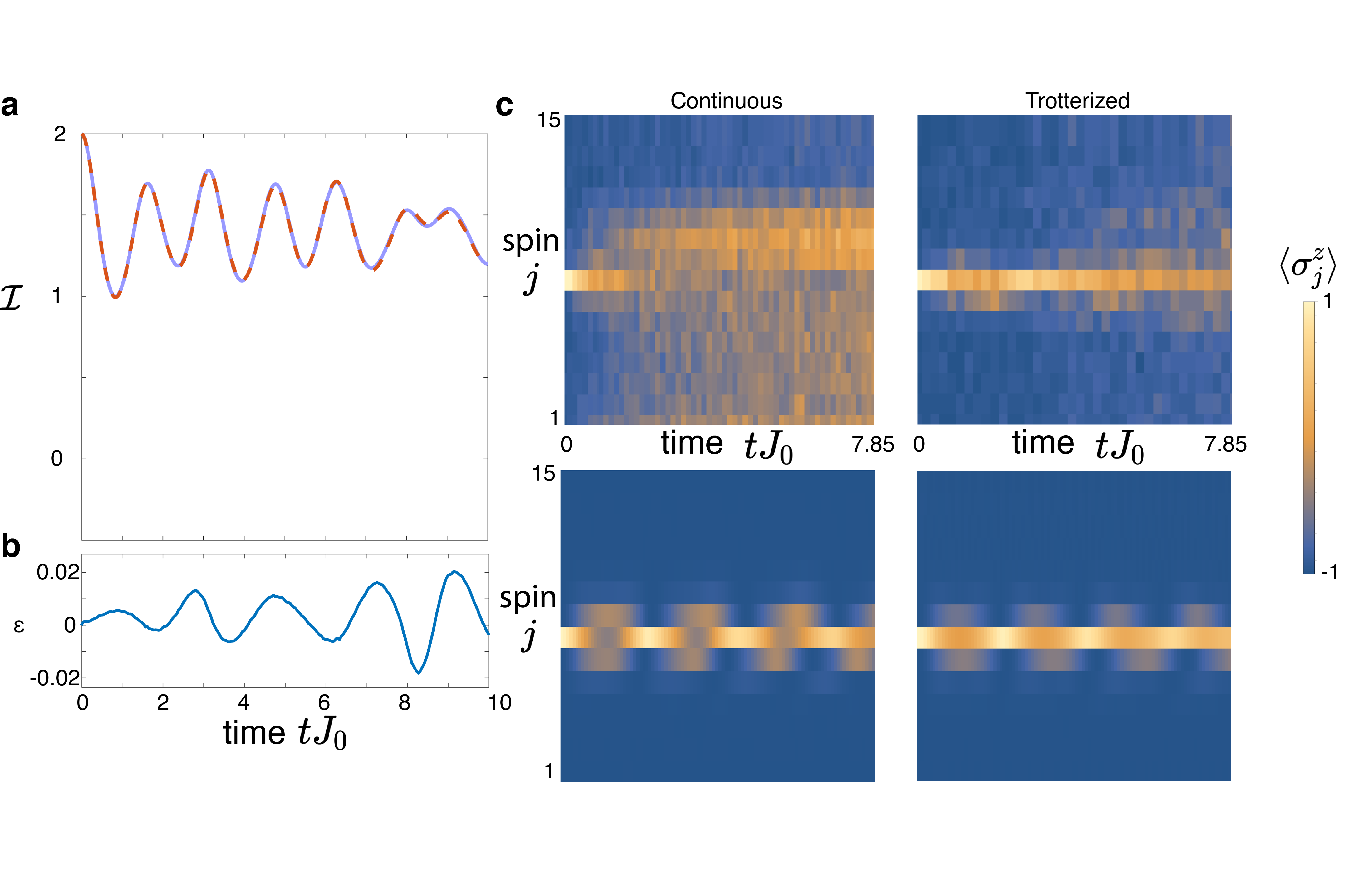}
\caption{Trotterization scheme. \textbf{a}, Numerics comparison of the imbalance dynamics for the averaged Hamiltonian of Eq.~\ref{eq:effHam} (solid blue line) with the full Trotter evolution (dashed orange), for the case of an initial N\'eel state $(N=15)$ and parameters corresponding to the strongest experimental field gradient. \textbf{b}, Difference (averaged - Trotter) between the plots in \textbf{a}, showing that the Trotter error over experimental timescales is on the order of one percent. \textbf{c}, Experimental examples (top row) of continuous and Trotterized evolution, both at $g/J_0=1.5$, compared to simulations (bottom row) using the (slightly different) parameters of the individual experimental realizations. Although the Trotterized evolution lasts nearly twice as much time in absolute units, since the averaged $J_0$ is roughly half as large, it nonetheless shows a substantial reduction in decoherence and improvement in fidelity to the desired Hamiltonian. An initial state with one spin flip is chosen for this comparison, as it makes the effect of decoherence due to phonons more pronounced compared with a state near zero net magnetization.}
\label{fig:FigTrotter}
\end{figure*}

We generate two types of Hamiltonian terms in this work. The first is the M\o{}lmer-S\o{}rensen Hamiltonian in the resolved sideband and Lamb-Dicke limits \cite{Monroe2019a}, created with a pair of detuned bichromatic beatnotes:
\begin{align}
    H_1(t) &= \sum_{j,\nu}\sigma_j^+\left [\frac{-i\Omega \eta_\nu b^\nu_j}{2} (a_\nu e^{-i\omega_\nu t} + a^\dagger_\nu e^{i\omega_\nu t}) \right. \nonumber \\
    &\left. (e^{-i\delta_B t}-e^{-i\delta_R t})\right] +h.c.
\end{align}
Here $j$ is the ion index and $\nu$ is the normal mode index, $a_\nu$ is the destruction operator of a phonon of motion for a given normal mode of the ion chain, $\Omega$ is the carrier Rabi rate, $\eta_{\nu}$ is the Lamb-Dicke parameter, $b^\nu_j$ is the mode amplitude for ion $j$, $\omega_\nu$ is the mode frequency, and $\delta_{B(R)}$ is the blue(red) detuning. This term generates spin-motion entanglement, and in the limit $\eta_{\nu} \Omega \ll |\delta_{R,B}-\omega_\nu|$ the motion can be adiabatically eliminated for an effective spin-spin interaction.

The second Hamiltonian term is the local field generated by the individual addressing beam. This beam only addresses one ion at a time, and is rastered across the chain to create an overall field landscape. A single cycle of this term can be written as:
\begin{equation}
    H_2(t)=\sum_j^N B^z_j\sigma_j^z \Theta(t-(j-1)t_{\text{pulse}})\Theta(jt_{\text{pulse}}-t),
\end{equation}
with $\Theta(t)$ as the Heaviside theta and $t_{\text{pulse}}$ the time for a pulse of the beam on one ion, which we experimentally fix at $t_{\text{pulse}}=0.5$ $\mu$s.

When these terms are applied simultaneously, in the limit $|\delta_{R,B}-\omega_\nu|\gg \eta_{\nu} \Omega \gg B^z_j$, the transverse Ising Hamiltonian is approximately realized:
\begin{equation}
    H_{TFIM}=\sum_{j,j'} J_{jj'} \sigma^x_j\sigma^x_{j'}+\sum_{j}\frac{B^z_j}{N}\sigma^z_j.
\end{equation}
However, the validity of this Hamiltonian is limited to small $B^z_j$. Therefore, when realizing a linear field gradient, $B^z_j=gNj$, this results in the constraint $gN^2\ll\eta_\nu \Omega$, which prevents the simultaneous attainment of long chains and large linear field gradients. For example, for typical experimental parameters of $N=15$, $\eta \Omega=2\pi \cdot 30$ kHz, and $J_0=2\pi \cdot$ 250 Hz, this would require that $g/J_0\ll0.5$. When this is not satisfied, additional phonon terms are present in the Hamiltonian that result in undesired spin-motion entanglement, or effective decoherence of the dynamics when measuring only spin.

We can reduce these constraints by applying a Trotterized Hamiltonian \cite{Lanyon2011,Zhu2021}. The evolution under this time-varying Hamiltonian can be analyzed using the Magnus expansion, to find the dominant contributions to time-averaged dynamics \cite{Monroe2019a}. Within this framework, the undesired effects arise from the commutator $[H_1(t),H_2(t)]$. Intuitively, when these terms are no longer applied simultaneously the effect of this commutator is reduced.

Consider unitary evolution of a single Trotter cycle, using the lowest-order symmetrized sequence:
\begin{align}
 U&=e^{-i \int_0^{\Delta t_2/2} H_2(t) dt} \nonumber \\
 &\times e^{-i \int_{\Delta t_2/2}^{\Delta t_1+\Delta t_2/2} H_1(t) dt} e^{-i \int_{\Delta t_1+\Delta t_2/2}^{\Delta t_1+\Delta t_2} H_2(t) dt}
 \label{eq:Trotter}
\end{align}
The Hamiltonians governing each part of the unitary evolution may be approximately replaced by their time-averaged values, simplifying both. For $H_2$ we have 
\begin{align}
    \int_0^{\Delta t_2/2} H_2(t) dt&=\nonumber \\
    \int_0^{\Delta t_2/2} \sum_j B^z_j\sigma_j^z &\Theta(t-(j-1)t_{\text{pulse}})\Theta(jt_{\text{pulse}}-t) dt \nonumber \\
    &=\frac{\Delta t_2}{2N}\sum_j B^z_j\sigma_j^z,
\end{align}
an exact identity since each of the terms in $H_2(t)$ commute with one another. For $H_1(t)$ we have
\begin{align}
    &\int_0^{\Delta t_1}  dt \sum_{j,\nu}\sigma_j^+\left [\frac{-i\Omega \eta_\nu b^\nu_j}{2} (a_\nu e^{-i\omega_\nu t} + a^\dagger_\nu e^{i\omega_\nu t}) \right. \nonumber \\
    & \left. (e^{-i\delta_B t}-e^{-i\delta_R t})\right] +h.c.
\end{align}
However, this is just the usual $M-S$ Hamiltonian, and in the limit that $|\delta_{R,B}-\omega_\nu|t\gg1$ the only significant contributing terms are the stationary ones. When $\delta_R=-\delta_B$ this results in the pure $\sigma^x\sigma^x$ interaction. When instead a small rotating frame transformation is applied we generate the Ising Hamiltonian with a small overall transverse field \cite{Monroe2019a}:
\begin{equation}
    \int_0^{\Delta t_1}  dt H_1(t)\approx \Delta t_1\left( \sum_{j,j'} J_{jj'} \sigma^x_j\sigma^x_{j'}+ B^{z0} \sum_{j}\sigma^z_j \right).
    \label{eq:MSApprox}
\end{equation}

The combined evolution of the full Trotter cycle is then, to lowest order, described by the Hamiltonian
\begin{align}
    H&=\frac{\Delta t_1}{\Delta t_1+\Delta t_2}\sum_{j,j'} J_{jj'} \sigma^x_j\sigma^x_{j'}\nonumber\\
    &+ \sum_{j}\sigma^z_j\left(B^{z0} + \frac{\Delta t_2}{\Delta t_1+\Delta t_2}\frac{B^z_j}{N}\right)+\mathcal{O}(\Delta t^3).
    \label{eq:effHam}
\end{align}
We program $B^z_j$ to the desired functional form and absorb the factors with $\Delta t_1$ and $\Delta t_2$ into re-definitions of $J_0$ and $g$ or $\gamma$, leading to Eqs. 1 and 4 of the main text. The constant term $B^{z0}$ does not depend on these times, because it is created by moving into a rotating frame that is applied to the entire time evolution. This approximation requires that $|\delta_{R,B}-\omega_\nu|\Delta t_1\gg1$ (for Eq.~\ref{eq:MSApprox}), which is satisfied in the experiment: $|\delta_{R,B}-\omega_\nu|_{min}=\mu=2\pi\cdot 200$ kHz and $\Delta t_1 \geq 18$ $\mu$s, whose product is 22.6. Additionally, $\Delta t_1$ and $\Delta t_2$ must not be so long that the Trotter approximation (Eq.~\ref{eq:effHam}) breaks down. However, the low energy scale of $J_0$ and the use of the symmetrized Trotter form make this limit less constraining than the limit for continuous evolution, allowing us to reach $g/J_0=2.5$ (1.5) for 15 (25) spins. Because the Trotter error consists of undesired spin terms, rather than spin-phonon terms, it can also be easily simulated numerically. Extended Data Fig.~\ref{fig:FigTrotter} shows comparisons of the Trotterized and ideal evolution in the case of the strongest gradient, showing that the Trotter error is negligible over the experimental timescale and that the Trotterization results in a significant improvement in the simulation fidelity.

In addition to reducing phonon errors, this scheme has the advantage of allowing us to tune the average Hamiltonian (Eq.~\ref{eq:effHam}) simply by varying $\Delta t_1$ and $\Delta t_2$, because $[g/J_0]_{avg}=(\Delta t_2/\Delta t_1)g/J_0$. This capability allows us to scan over a range of gradient values with a single calibration, and it makes any errors on the gradient calibration common to all these scans. In the data presented here, we fix the instantaneous values of $g$ and $J_0$ and vary $\Delta t_1$ (see subsequent section, `Trotterized Hamiltonian parameters'). In addition, we ramp the spin-spin interactions up and down over 9 $\mu$s with a shaped Tukey profile to reduce adiabatic creation of phonons \cite{Zhang2017a}.

This implementation of Trotterized Stark MBL dynamics would be difficult to extend to more than tens of spins, as the maximum instantaneous shift required on the edge ion scales as $N^2$, leading to the requirement of an increasingly fast drive. However, given the unbounded nature of a linear gradient, any large- scale simulation of Stark MBL is likely to be challenged by the required field difference between the two ends.

Throughout this discussion, we have taken the perspective of a Trotterized quantum simulation of a desired Hamiltonian. We could also understand this experiment in terms of Floquet theory. From this perspective, this driven system is described stroboscopically by a Floquet Hamiltonian, which to lowest order is the Hamiltonian (\ref{eq:effHam}), and the steady-state equilibration that we see represents prethermal evolution under this effective Hamiltonian that is expected to be altered at long times by Floquet heating arising from the higher-order terms. While this picture offers a complementary way to understand these results, and interesting connections to studies of driven localization \cite{Ponte2015}, for simplicity we focus on the Trotterized perspective.

\subsection*{Trotterized Hamiltonian parameters}

For imbalance measurements at $N=15$, we calibrate to $g/J_0$ of 2.5 for $\Delta t_1=\Delta t_2$. To scan the gradient strength, $\Delta t_2$ is fixed at 18 $\mu$s and $\Delta t_1$ is varied from 18 $\mu$s to 180 $\mu$s. In addition, there is an extra 9 $\mu$s of effective dead time per Trotter step associated with the Tukey pulse shaping. We fix $B^{z0}$ at $2\pi \cdot$ 1.25 kHz. For data in a quadratic field, we set $\gamma=2.0$ for $\Delta t_1=\Delta t_2$, and vary $\Delta t_2$ from 10 $\mu$s to 180 $\mu$s, with all other settings kept the same as in the linear gradient.

For $N=25$, we instead set $g/J_0$ to 1.25 for $\Delta t_1=\Delta t_2$. $\Delta t_1$ is fixed at 30 $\mu$s, and $\Delta t_2$ is varied between 25 $\mu$s and 190 $\mu$s, again with an extra 9 $\mu$s of effective dead time per cycle due to pulse shaping. $B^{z0}$ is again fixed at $2\pi \cdot$ 1.25 kHz.

For DEER measurements, we calibrate to $g/J_0$ of 2.0. $\Delta t_2$ is fixed at 18 $\mu$s and $\Delta t_1$ is varied from 18 $\mu$s to 180 $\mu$s, plus an extra 9 $\mu$s of dead time associated with Tukey pulse shaping. We fix $B^{z0}$ at values varying for different datasets between $2\pi \cdot$ 0.9 kHz and $2\pi \cdot$ 1.25 kHz.

\section*{Analysis of the Hamiltonian}

\subsection*{Mapping to boson model}

Our experimental Hamiltonian, from Eq.~\ref{eq:ExpHam} of the main text, is:
\begin{equation}
H=\sum_{j<j'} J_{jj'} \sigma_{j}^{x} \sigma_{j'}^{x}+\sum_{j=1}^N (B^{z0}+(j-1)g) \sigma_{j}^{z}.
\label{eq:ExpHam2}
\end{equation}
In the limit of $B^{z0} \gg J_0$, and assuming that $B^{z0}$ and $g$ have the same sign, the total magnetization $\sum_{j}\left\langle\sigma_{j}^{z}\right\rangle$ is conserved. For an initial state of definite total magnetization, the system then reduces to the long-range tilted XY Hamiltonian \cite{Richerme2014}:
\begin{align}
    H_{XY}&=\sum_{j<j^{\prime}} \frac{J_{j j^{\prime}}}{2}\left(\sigma_{j}^{+} \sigma_{j^{\prime}}^{-}+\sigma_{j}^{-} \sigma_{j^{\prime}}^{+}\right) \nonumber \\
    &+\sum_{j=1}^{N}(B^{z0}+(j-1)g)\sigma^z_j.
    \label{longXY}
\end{align}
This can be mapped to a system of hard-core bosons taking $\sigma^{-(+)}_j \rightarrow a_j^{(\dagger)}$ and $n_j=a^\dagger_ja_j=(\sigma^z_j+1)/2$, resulting in the Hamiltonian:
\begin{align}
    H_{HC}&=\sum_{j<j^{\prime}} \frac{J_{j j^{\prime}}}{2}\left(a_{j}^{\dagger} a_{j^{\prime}}+a_{j} a_{j^{\prime}}^{\dagger}\right)+U\sum_{j=1}^N n_j (n_j-1)\nonumber\\&+ \sum_{j=1}^{N}(\mu_N+2(j-1) g) n_j,
\end{align}
with $\mu_N=2B^{z0}$, taking the limit $U\rightarrow \infty$, and dropping a constant energy contribution.

This model clarifies the connection between our system and work studying Stark MBL in the context of hopping particles with interactions \cite{VanNieuwenburg2019,Schulz2019}. It also illustrates the translational symmetry in our system. If $j$ is shifted by an integer, this is equivalent to changing the chemical potential term $\sum_j \mu_N n_j$, which has no effect in a closed system with particle conservation.

\subsection*{Gauge transformation of the Hamiltonian}

The linear potential in this model can be removed using a gauge transformation $\mathcal{U}$ \cite{VanNieuwenburg2019,Scherg2020}:
\begin{equation}
    \mathcal{U}=e^{it\sum_j (\mu_N+2(j-1) g) n_j }.
\end{equation}
After this transformation, which is equivalent to moving into the interaction picture with respect to the local field term, the transformed Hamiltonian is:

\begin{align}
    H_{HC}'(t)&=\sum_{j<j^{\prime}} \frac{J_{j j^{\prime}}}{2}\left(a_{j}^{\dagger} a_{j^{\prime}}e^{-2ig(j-j')t}+a_{j} a_{j^{\prime}}^{\dagger}e^{2ig(j-j')t}\right) \nonumber\\
    &+U\sum_{j=1}^N n_j (n_j-1).
\end{align}
In the limit of short-range terms, the time dependence of this transformed Hamiltonian has a bounded set of frequencies, and going to the thermodynamic limit is straightforward. However, long-range terms result in time dependence that becomes arbitrarily fast for terms with arbitrarily large $|j-j'|$. This points to a fundamental difference between short- and long-range Hamiltonians in the presence of superextensive potential terms.

\subsection*{Effective Hamiltonian from Schrieffer-Wolff transformation}

 To understand the mechanism of Stark MBL, it is useful to derive the effective Hamiltonian in the limit of a strong tilt. To do this, we apply degenerate perturbation theory in the small parameter $J_0/g$ to Eq.~(\ref{longXY}), in a variation of the Schrieffer-Wolff transformation \cite{Schrieffer1966,Yang2020}. The goal is to construct a unitary transformation:
\begin{align}
\label{eq:SWdef}
H_{\rm eff}&=e^S\, H\, e^{-S} \nonumber \\
&= H+[S,H]+\frac{1}{2!}[S,[S,H]]+\frac{1}{3!}[S,[S,[S,H]]]+\dots \nonumber \\
&=\sum^{\infty}_{n=0}H^{(n)}_{\rm eff}.
\end{align}
Here we have the Schrieffer-Wolff generator, $S$, which is anti-Hermitian, and $H^{(n)}_{\rm eff}$ of order $(J_0/g)^n$.  The form for $S$ is determined by separating the Hamiltonian into diagonal and off-diagonal contributions in the $\sigma^z_j$ basis:
\begin{align}
H&=H_0+V,\\
H_0&= \sum_{j=1}^{N}(B^{z0}+(j-1)g)\sigma^z_j,  \\
V&= \sum_{j<j'} \frac{J_{j j^{\prime}}}{2}\left(\sigma_{j}^{+} \sigma_{j^{\prime}}^{-}+\sigma_{j}^{-} \sigma_{j^{\prime}}^{+}\right).
\end{align}
Then, $S$ is chosen to eliminate block-off-diagonal contributions to $H_{\rm eff}$ at each order, leading to the condition that $[H^{(n)}_{\rm eff},H_0]=0$ for each $n$. This enforces center-of-mass (or dipole moment) conservation at each order. As a result, $S$ has the following form:
\begin{align}
S = \sum^{\infty}_{n=1}S^{(n)},
\end{align}
with $S^{(n)}$ of order $(J_0/g)^n$. Applying this form to Eq.~\ref{eq:SWdef} and organizing the terms by powers of $J_0/g$ results in:

\begin{widetext}

\begin{align}
\label{eq:SWexpansion}
H_{\rm eff}&=H_0\!+\!\left([S^{(1)}\!,H_0]\!+\!V\right)\!+\!\left([S^{(2)}\!,H_0]\!+\![S^{(1)}\!,V]\!+\!\frac{1}{2!}[S^{(1)}\!,[S^{(1)}\!,H_0]]\right) \nonumber \\
\quad\!&+\!\left([S^{(3)}\!,H_0]\!+\![S^{(2)}\!,V]\!+\!\frac{1}{2!}\left([S^{(1)}\!,[S^{(1)}\!,V]]\!+\![S^{(1)}\!,[S^{(2)}\!,H_0]]\!+\![S^{(2)}\!,[S^{(1)}\!,H_0]]\right)\!+\!\frac{1}{3!}[S^{(1)}\!,[S^{(1)}\!,[S^{(1)}\!,H_0]]]\right)\!+\!\dots.
\end{align}

\end{widetext}
With this form, $S^{(n)}$ must then be chosen to make $[S^{(n)},H_0]$ cancel all block-off-diagonal (i.e., non-dipole-conserving) terms at order $n$. While the resulting expression is inconvenient to write out explicitly, this approach can be applied algorithmically to find arbitrarily high orders.

Alternatively, one may set $S^{(n)}=0$ for all $n\geq 2$ and manually project out non-dipole-conserving terms order by order. $S^{(1)}$ must still obey the constraint $[S^{(1)},H_0]+V=0$, which can be achieved by taking the form
\begin{align}
\label{eq:S1def}
\langle\sigma|S^{(1)}|\sigma^\prime\rangle
=\frac{\langle\sigma|V|\sigma^\prime\rangle}{\langle\sigma|H_0|\sigma\rangle-\langle\sigma^\prime|H_0|\sigma^\prime\rangle}.
\end{align}
The resulting leading-order effective Hamiltonian is
\begin{align}
\label{eq:HeffSW}
H_3^{eff}=& \sum_{i<j<k<l; i+k= j+l} \frac{3( \sigma_i^{+}\sigma_{j}^{-} \sigma_{k}^{-} \sigma_{l}^{+} +H.c.)}{(j-i)(k-i) g^2} \nonumber \\
&*(J_{ij} J_{jk} J_{jl} +J_{ik} J_{jk} J_{kl} - J_{ij} J_{ik} J_{il} - J_{il} J_{jl} J_{kl})
\end{align}
(where we have omitted lower-order energy correction terms that are diagonal in the $H_0$ basis). Starting from an initial state that is an eigenstate of $H_0$, the effective Hamiltonian couples this state to other eigenstates of $H_0$ with the same energy. This directly translates to the dipole conservation constraint $i+k=j+l$ in Eq.~(\ref{eq:HeffSW}). 
Although the above process comes from the third-order contribution to $H^{eff}$, the effective Hamiltonian contains only four-body terms that conserve the dipole moment. 
Note that the above effective Hamiltonian does not vanish even for translationally invariant long-range couplings. For the case of long-range couplings that can approximated by power-law decay $J_{i-j}= J_0/|i-j|^{\alpha}$, the above equation can be written as
\begin{align}
H_3^{eff}= \sum_{i<j<k<l; i+k= j+l} & \frac{6 J_0^3( \sigma_i^{+}\sigma_{j}^{-} \sigma_{k}^{-} \sigma_{l}^{+} +H.c.)}{g^2 (j-i)^{\alpha+1}(k-i)^{\alpha+1} } \nonumber \\
& \times \left[\frac{1}{(k-j)^{\alpha}}-\frac{1}{(l-i)^{\alpha}} \right]
 \label{eq:SWEffectiveHamiltonian}
\end{align}
This is in contrast with a short-range XY Hamiltonian with nearest-neighbor interactions, where the above term vanishes in the limit of $\alpha=\infty$.

The effective Hamiltonian Eq. \ref{eq:SWEffectiveHamiltonian} shows that dipole-conserving terms with arbitrarily long range exist in this system even in the lowest nontrivial order of the perturbative expansion. The strengths of these long-range coupling terms decrease monotonically with the power-law exponent $\alpha$. This result can be contrasted with two other cases. A short-range Hamiltonian with dipole and spin-flip conservation can result in Hilbert space fragmentation (or shattering) \cite{Schulz2019,VanNieuwenburg2019,Khemani2020,Moudgalya2019,Sala2020,Taylor2019}, while in this long-range model fragmentation is not present in the thermodynamic limit \cite{Sala2020}. On the other hand, a similar perturbative expansion beginning with a short-range tilted model will also give long-range dipole-conserving terms, but only at higher powers of the tilt \cite{Moudgalya2019,Scherg2020}.


Despite the lack of fragmentation, this Hamiltonian does result in state-dependent relaxation. One reason for this is that the dipole conservation term in Eq.~(\ref{eq:HeffSW}) depends on the  distances between the four operators. Specifically, $J_{ij}$ decays as a function of distance between the pair of ions. Additionally, the denominator in the above expression contains the factors $(j-i)$ and $(k-i)$. The distance dependence comes from the energy differences between intermediate states of the perturbation theory. This combination of distance-dependent factors can result in different slow delocalization dynamics for the different initial states shown in Fig.~\ref{fig:Fig2} (see Methods section `Long-term stability of Stark MBL').




\section*{Full level statistics of experimental Hamiltonian}

\begin{figure}
\centering
\includegraphics[width=0.45\textwidth]{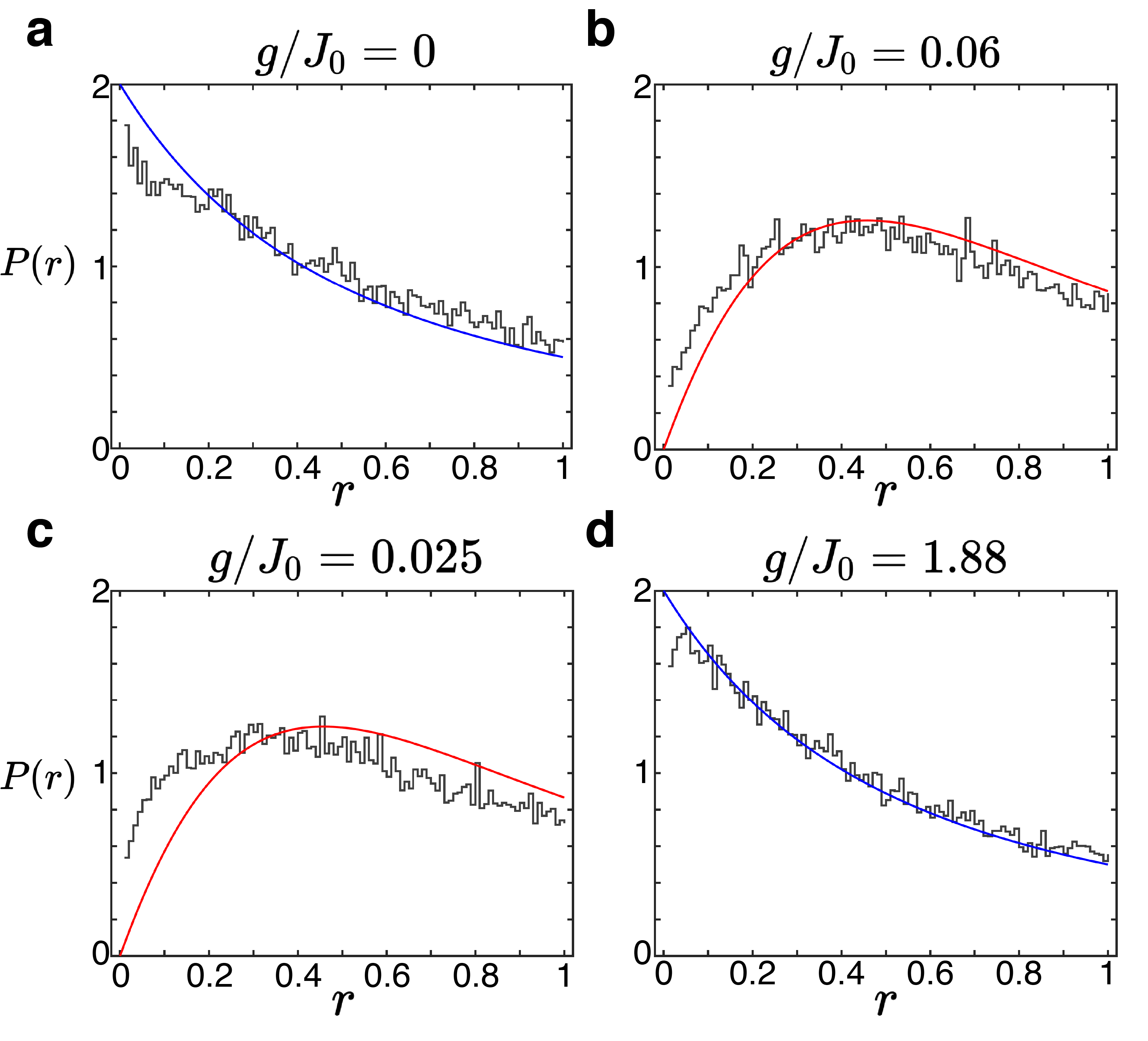}
\caption{Probability density distributions of $r$, the ratio of adjacent energy level spacings, for the experimental Hamiltonian (Eq.~1 of the main text) at various values of $g/J_0$ and $N=15$. Numerics are compared with the distribution expected for either a Poisson level distribution (blue lines in \textbf{a} and \textbf{d}) or a Wigner-Dyson distribution (red lines in \textbf{b} and \textbf{c}). The level statistics in the absence of a field gradient are near the Poissonian limit, which may reflect the proximity to an integrable limit for the low-energy sector \cite{Neyenhuis2017}. A small gradient results in statistics near the Wigner-Dyson limit, followed by an approach to Poisson statistics as the gradient is increased.}
\label{fig:FigHistogram}
\end{figure}

 A typical ergodic system has a reduced single-particle density matrix with support throughout the bulk, and thus has a high degree of overlap between particles. This results in level repulsion in the many-body spectrum, leading to a Wigner-Dyson energy level distribution characteristic of random matrices \cite{Oganesyan2007}. A typical localized system, on the other hand, has single particles that are spatially confined, and thus have little overlap, resulting in a Poissonian distribution of the many-body spectrum. In Extended Data Fig.~\ref{fig:FigHistogram} we show the full distribution of $r$, the ratio of adjacent energy level spacings, for the experimental Hamiltonian at selected values of $g/J_0$. We compare it to the probability density distributions resulting from Poisson and Wigner-Dyson statistics \cite{Schulz2019}:
\begin{align}
    P_p(r) &= \frac{2}{(1+r)^2}\text{ (Poisson),}\\
    P_{WD}(r) &= \frac{27(r+r^2)}{4(1+r+r^2)^{5/2}}\text{ (Wigner-Dyson),}
    \label{eq:WDeq}
\end{align}
where Eq.~\ref{eq:WDeq} is an analytic approximation to the Gaussian Orthogonal Ensemble based on the Wigner Surmise \cite{Atas2013}.

While a small field gradient is needed to break the approximate integrability of the Hamiltonian \cite{Neyenhuis2017} in the limits of $g=0$ and $B^{z0}\gg J_0$, over the range of tilts studied experimentally the level statistics cross from being close to the Wigner-Dyson limit, with an evident dip at low $r$ due to the proliferation of avoided crossings, to very close to the Poisson limit at large gradients. This should be contrasted with the case of short-range hopping, in which the level statistics may be highly non-generic due to exact degeneracies associated with dipole conservation, making concepts of Hilbert space fragmentation (or shattering) especially relevant \cite{Schulz2019,Taylor2019,VanNieuwenburg2019,Scherg2020,Sala2020,Khemani2020,Moudgalya2019,Yang2020,Li2021}. Although the level statistics shown here are for an experimentally measured Hamiltonian, featuring small deviations from a perfectly linear gradient, these deviations do not substantially affect the level statistics, as the long-range terms already lift the degeneracies. In the next section we show this explicitly, using the ideal power-law Hamiltonian to study more general features of Stark MBL with long-range couplings such as the scaling behavior.

\section*{Numerical studies of the ideal power-law Hamiltonian}

The experimental system is approximately described by a Hamiltonian with a power-law hopping:
\begin{equation}
H=\sum_{j<j'} \frac{J_0}{|j-j'|^{\alpha}} \sigma_{j}^{x} \sigma_{j'}^{x}+\sum_{j=1}^N (B^{z0}+(j-1)g) \sigma_{j}^{z}.
\label{eq:ExpHamPowerLaw}
\end{equation}
However, as the exact experimental couplings feature inhomogeneity across the chain and deviations from power-law scaling for large ion separations, all numerics shown in the main text (as well as the previous sections) use the exact Hamiltonian as determined by experimental measurements of mode structure and detuning. Nonetheless, to study the general behavior of the system it is useful to also look at the power-law Hamiltonian, which captures the dominant behavior while being translation-invariant and therefore having a more natural scaling with size. We study this numerically to characterize the behavior of $\langle r \rangle$ with respect to $\alpha$ and $g/J_0$, and to study the finite-size dependence.

\subsection*{Dependence of $\langle r \rangle$ on $\alpha$ and $g/J_0$}

\begin{figure}
\centering
\includegraphics[width=0.45\textwidth]{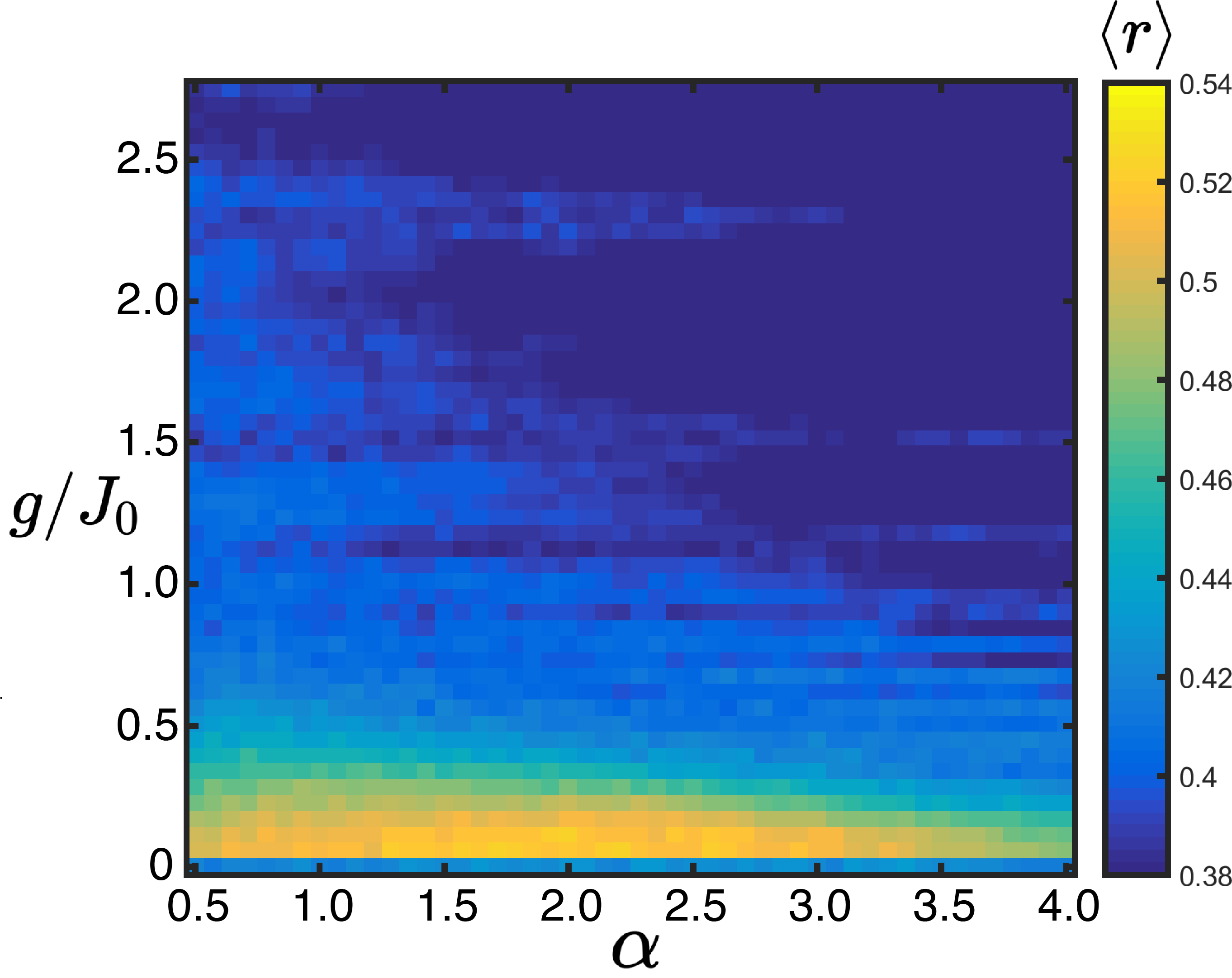}
\caption{Dependence of $\langle r \rangle$ on power-law range $\alpha$ and $g/J_0$ (N=13, $B^{z0}/J_0=5$). In the experiments presented in the main text $\alpha \approx 1.3$.}
\label{fig:FigPD}
\end{figure}

Extended Data Fig.~\ref{fig:FigPD} shows the dependence of the level statistics $\langle r \rangle$ on the Hamiltonian parameters $\alpha$ and $g/J_0$. The primary features of the experimental Hamiltonian statistics are retained, such as non-generic statistics for very small gradient values and a crossover from $\langle r \rangle \approx 0.5$ to 0.39 for $g/J_0$ between 0.1 and 2.0. For $\alpha<1$, the concept of Stark MBL may break down entirely, as the spin-spin coupling energy is superextensive. While we see some signature of this in Extended Data Fig.~\ref{fig:FigPD}, such as the increase in the gradient needed to reach Poissonian statistics as $\alpha$ is decreased, near $\alpha=1$ the divergence of the spin-spin energy with system size is logarithmically slow, making finite-size effects significant.

For large $\alpha$, $\langle r \rangle$ generally decreases, which reflects the approach to the limit of Wannier-Stark localization because the short-range model maps to a chain of free fermions with a tilt under a Jordan-Wigner transformation. The general features observed are consistent with a recent study of long-range hopping in a tilt \cite{Bhakuni2020} that also found persistence of a crossover in $\langle r \rangle$ up to $N=18$ and for $\alpha>1$.

\subsection*{Dependence of $\langle r \rangle$ on system size}

\begin{figure}
\centering
\includegraphics[width=0.4\textwidth]{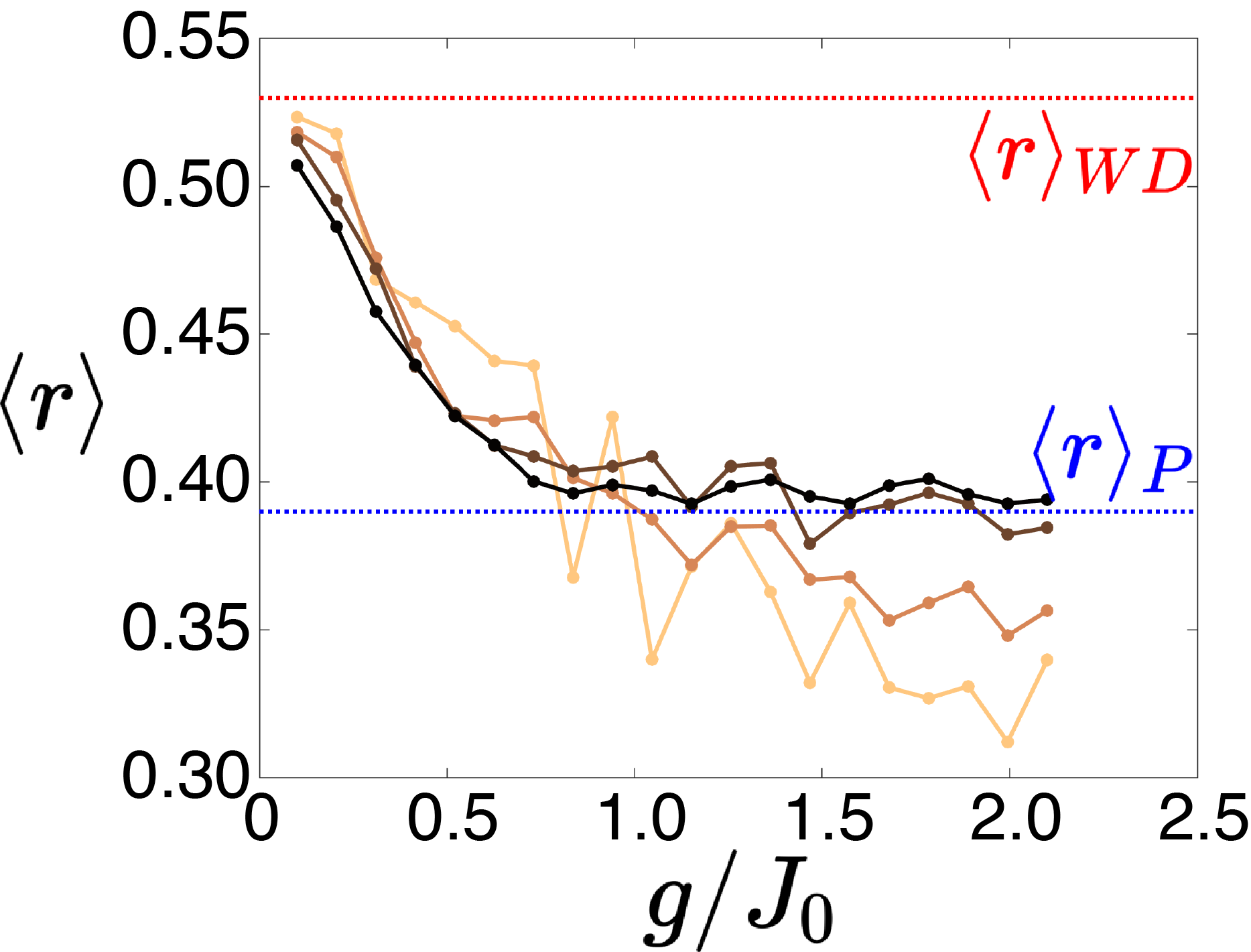}
\caption{Dependence of level statistics on system size for the power-law Hamiltonian (Eq. \ref{eq:ExpHamPowerLaw}). Level statistics for $N=$ \{9,11,13,15\} (light to dark), for $\alpha=1.3$ and $B^{z0}/J_0=5$.}
\label{fig:FigScaling}
\end{figure}

Using the power-law Hamiltonian, we can study the dependence of the level statistics on system size. Extended Data Fig.~\ref{fig:FigScaling} shows this for $N$ ranging from 9 to 15. In general, the curves do not exhibit a simple finite-size scaling. This may be due to the long-range couplings, which are known to cause a system size-dependent shift in the transition in numerics for the disordered MBL case \cite{Wu2016}. The progressive shift away from the Wigner-Dyson limit at small gradient may indicate that this regime is `quasi-ergodic' due to finite-size effects \cite{Doggen2020}, or reflect anomalous thermalization \cite{Gromov2020}, or may instead reflect an increasing effect of the non-generic statistics observed near zero gradient in the previous section (`Dependence of $\langle r \rangle$ on $\alpha$ and $g/J_0$'). Crucially, we see that the trend of gradient-driven localization persists up to the largest systems we can diagonalize, coinciding with the size used for most of the data presented in the main text, with a full study of the scaling left as an interesting subject for future work.

\subsection*{Dependence of $\overline{\mathcal{I}}$ on system size}

\begin{figure}
\centering
\includegraphics[width= 0.45 \textwidth]{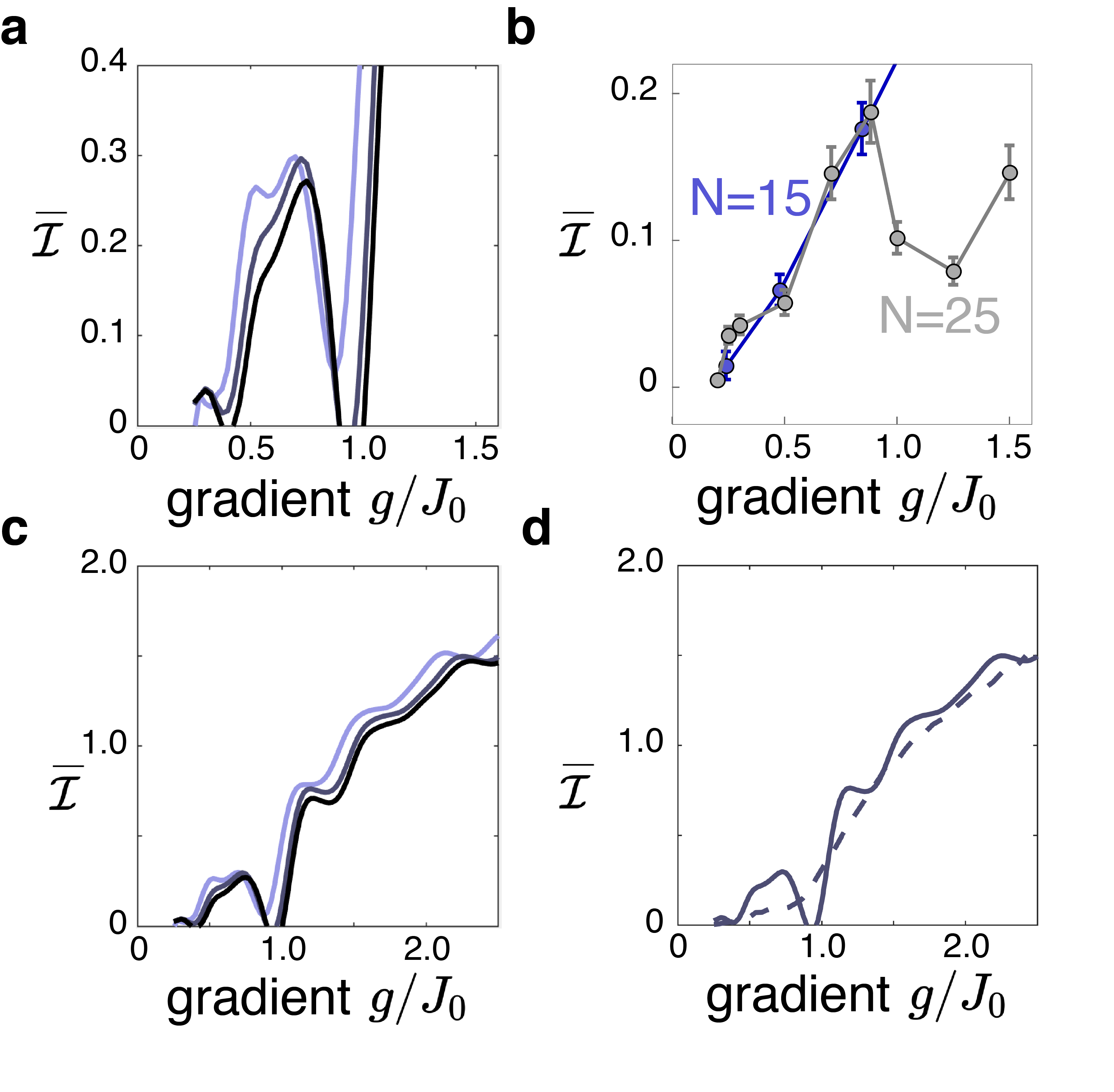}
\caption{Dependence of $\overline{\mathcal{I}}$ on system size and time. \textbf{a}, Numerics showing $\overline{\mathcal{I}}$ for the N\`eel state with $N=$\{9, 15, 25\} (light to dark). As the system increases from $N=9$ to $N=25$, the largest change is in a sharpening feature near $g/J_0=1$. These numerics do not include experimental noise. \textbf{b}, Experimental data for $N=15$ and $N=25$, reproduced from Fig.~\ref{fig:Fig2}, shows a similar dip for the larger size. \textbf{c}, Expanded view of numerics from \textbf{a}. Especially for gradient values above $g/J_0=1$, the imbalance shows little finite-size dependence. \textbf{d}, Numerical comparison of $\overline{\mathcal{I}}$ ($N=15$) for the experimental time and for an extended time of 100 $tJ_0$ (dashed). While at small gradients the finite-time effects on the imbalance are significant, including the dip feature in the left plots, a steady state is largely achieved in the experimental window for gradients $g/J_0>1$. For all numerics shown, $B^{z0}/J_0=4.4(1+3g/(5J_0))$ (the experimental scaling resulting from Eq.~\ref{eq:effHam} with $\Delta t_1$ varied)  and $\alpha=1.3$.}
\label{fig:FigFinizeSizeImbal}
\end{figure}

Extended Data Fig.~\ref{fig:FigFinizeSizeImbal} shows a comparison of our data for $\overline{\mathcal{I}}$ varying system size (Fig.~\ref{fig:Fig2}e) with numerics. We present data for $N=9$, $N=15$, and $N=25$, corresponding to size increases by a factor of $5/3$.

For the most part, $\overline{\mathcal{I}}$ only shows a slight shift with increasing $N$. However, there is a sharp feature near $g/J_0=1.0$ that grows more prominent with increasing size, and appears similar to the experimental dip observed for $N=25$. This feature is a finite-time effect, as seen in Extended Data Fig.~\ref{fig:FigFinizeSizeImbal}, and also depends on the initial state. It reflects the complex dynamical possibilities for $g/J_0<1$, in which various tunneling processes are energetically permitted. However, interpretation of this feature in experimental data is complicated by decoherence that increases both with $g/J_0$ and with $N$.

In general, these initial-state dependent dynamics for $g/J_0<1$ may display rich possibilities such as subdiffusion \cite{Guardado-Sanchez2020,Gromov2020}, complicating any determination of a critical transition value from quench dynamics \cite{Doggen2020}. However, for $g/J_0>1$ the transient dynamics are simpler, and the imbalance comes close to its long-lived steady-state value within the experimental window.

\subsection*{Long-time stability of Stark MBL}

\begin{figure*}
\centering
\includegraphics[width=0.95\textwidth]{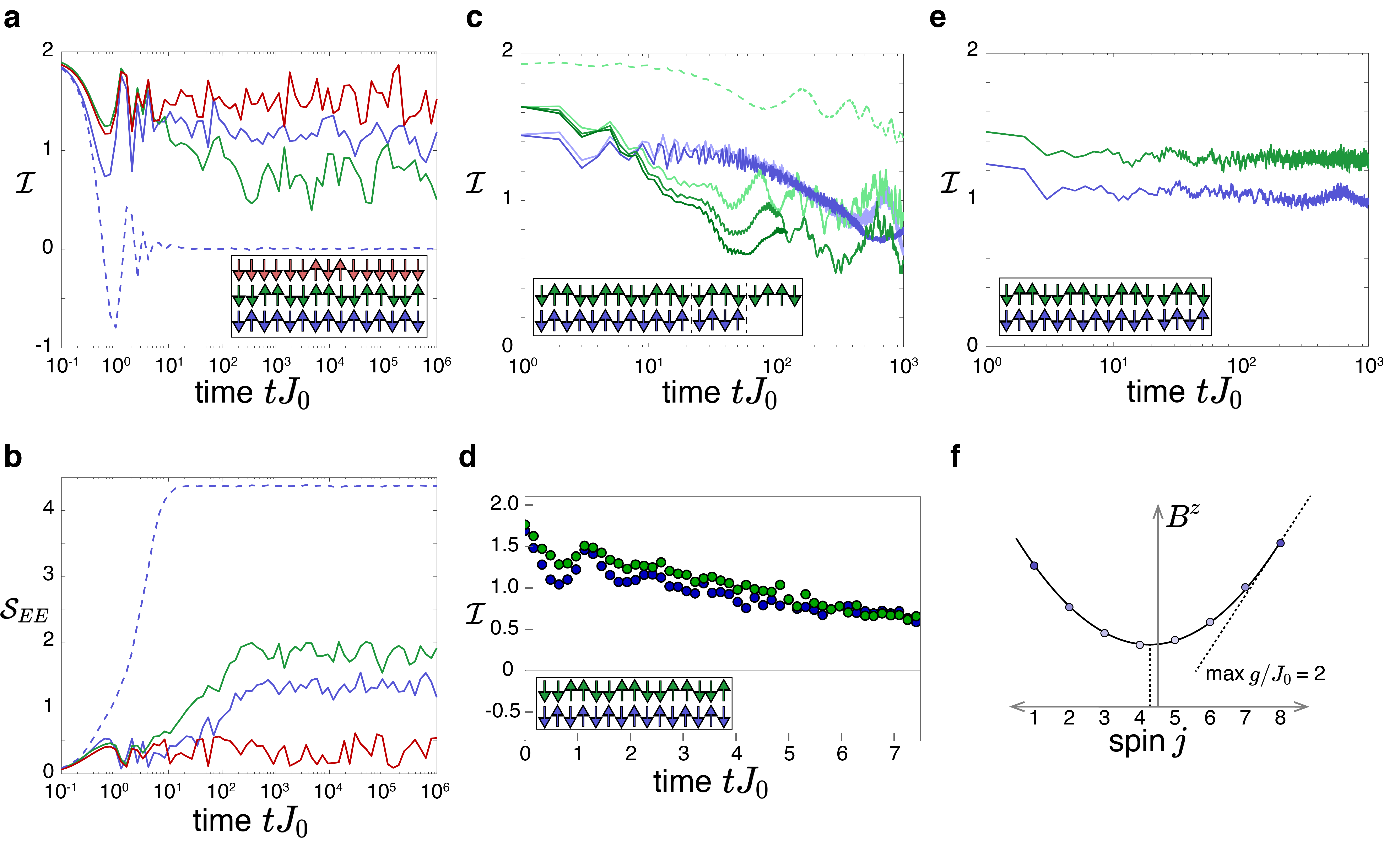}
\caption{Long-term stability of Stark MBL. \textbf{a,b}, Numerical study of the long-time dynamics of the initial states realized in Fig.~2, using exact diagonalization.
For this finite-size realization, in a strong gradient ($g/J_0=2$, solid lines), the imbalance and bipartite entanglement entropy show some slow dynamics but apparently never approach the thermal value, in contrast with a weak gradient ($g/J_0=0.25$, dashed line).
\textbf{c}, Numerical study of the finite-size and initial-state dependence of Stark MBL imbalance dynamics. States with one-block (N\`eel) and two-block domain walls are shown for $g/J_0=2$ and $N=12$, $N=16$, and $N=20$ (light to dark solid lines, $N=20$ for the two-block state only). The two-block initial state shows faster decay and greater finite-size effects, as is expected from the effective Hamiltonian in a large tilt [Eq.~(\ref{eq:HeffSW})]. With a stronger gradient (dashed line, $g/J_0=5$ and $N=12$), this instability can be arbitrarily postponed. To show the long-term trend clearly, a moving average with a window of 5$J_0$ has been applied to these numerics. \textbf{d}, Experimental data for the one and two-block domains. Consistent with numerics, state-dependent instability is manifested as a slow differential increase in the decay of the two-block state compared to the N\`eel state. These data were taken consecutively to ensure identical experimental parameters and decoherence rates. Each point is an average over 200 experimental repetitions, with error bars smaller than the symbol size. \textbf{e}, Numerical studies of stability in a quadratic field ($N=16$, $\gamma=2$) do not show this state-dependent instability over the same timescale. To show the long-term trend clearly, a moving average with a window of 5$J_0$ has been applied to these numerics. \textbf{f}, Cartoon of the setup for numerics in \textbf{e} (shown with $N=8$ for clarity). The quadratic potential is chosen to have a minimum shifted away from the system center by one-quarter site to avoid a fine-tuned reflection symmetry. For all numerics shown, $B^{z0}/J_0=4.5$ and $\alpha=1.3$.}
\label{fig:ExactDiag}
\end{figure*}

A subject of much debate in the study of localization is the stability of the localized state to various slow delocalization processes. In the context of Stark MBL, these might include coupling between many-body states with the same spin and dipole quantum numbers, or slow dipole-moment changing processes \cite{Moudgalya2019,Khemani2020,Doggen2020}. These questions are most relevant for the ideal power-law Hamiltonian, as such slow processes could conceivably be halted by even the small amount of residual disorder or inhomogeneity in our experimental realization. To study this possibility, Extended Data Fig.~\ref{fig:ExactDiag} shows the dynamics for very long times of the quenched initial states studied in Fig.~\ref{fig:Fig2}, using the ideal disorder-free power-law Hamiltonian.

We find several noteworthy results. First, in a finite system such as those realized in our experiment, some Stark MBL localization appears to persist indefinitely.  This is striking, as relaxation is not forbidden by energetics, nor by any other conservation law.

Second, in a finite-size numerical analysis, we see increasing amounts of slow, state-dependent relaxation, which may make Stark MBL unstable in the thermodynamic limit. This relaxation can be understood via the effective Hamiltonian [Eq.~(\ref{eq:HeffSW})] in the large-gradient limit. For the two-block state with the configuration $01100110011$ (where 0 and 1 represent down and up spins, respectively), the most significant contribution from this effective Hamiltonian is the process $1001 \leftrightarrow 0110$. This is also the largest term in the effective Hamiltonian, making the stability of this state the most restrictive condition for localization. However, for the N\`eel state with the configuration 0101010101, the most significant contribution is $01010 \leftrightarrow 10001$. Both processes appear at the same order of the Hamiltonian, but with different strengths. When $\alpha=1.3$, the process $1001 \leftrightarrow 0110$ has an amplitude of $0.96 J_0^3/g^2$, while the process $01010 \leftrightarrow 10001$ has an amplitude of 0.22 $J_0^3/g^2$. This explains in part why we see faster relaxation for the two-block state, although, as we are not deep in the $g\gg J_0$ limit, higher terms are expected to contribute as well. These observations are also consistent with previous work showing that for cases in which the effective Hamiltonian has multiple dipole-conserving terms with different ranges and strengths, thermalization can be very slow or absent entirely for finite-sized systems \cite{Taylor2019}. We emphasize that although state-dependent relaxation has been proposed as an experimental signature of (exact or approximate) Hilbert space fragmentation \cite{Khemani2020,Scherg2020}, we realize a similar phenomenon here, with less separation between the timescales of the different decay processes, without true fragmentation due to our long-range couplings. 

This state-dependent relaxation is evident experimentally as a small but robust state-dependent difference in the rate of decay of the imbalance. As a simple test, an exponential fit to the N\`eel state decay shown in Extended Data Fig.~\ref{fig:ExactDiag}d, excluding points before $tJ_0=2$, gives a time constant of $\tau_{Neel}J_0=8.6\pm0.46$, while the fit decay for the two-block state is $\tau_{2B}J_0=7.1\pm0.24$. These data sets were taken consecutively to avoid any experimental drift, and the differential decay shown is representative of other datasets at similar parameters.

Because this delocalization is highly dependent on the linear form of the Stark MBL gradient, which enforces approximate dipole conservation, we may expect very different behavior in a quadratic field. This is confirmed in the right panel of Fig.~\ref{fig:ExactDiag}. After the initial dynamics of order $t\sim1/J_0$, no additional relaxation is observed for either state. While higher-order processes may still lead to relaxation in the thermodynamic limit, for relatively small systems this localization appears quite robust.

Summarizing, Stark MBL appears to be a relevant concept under any of several conditions: first, for finite-sized systems, in which thermalization can be postponed seemingly indefinitely. Second, in arbitrarily large systems over timescales that are short compared to $(g^2/J_0^3)$ (or possibly longer in systems without native long-range terms). And finally, in systems which have more constraints than a linear field, such as a linear field with nonzero curvature \cite{Schulz2019,Chanda2020, Yao2021} or disorder \cite{VanNieuwenburg2019} (a small amount of which is present in our experimental realization).



\section*{Quantum Fisher information}

\begin{figure}
\centering
\includegraphics[width= 0.45 \textwidth]{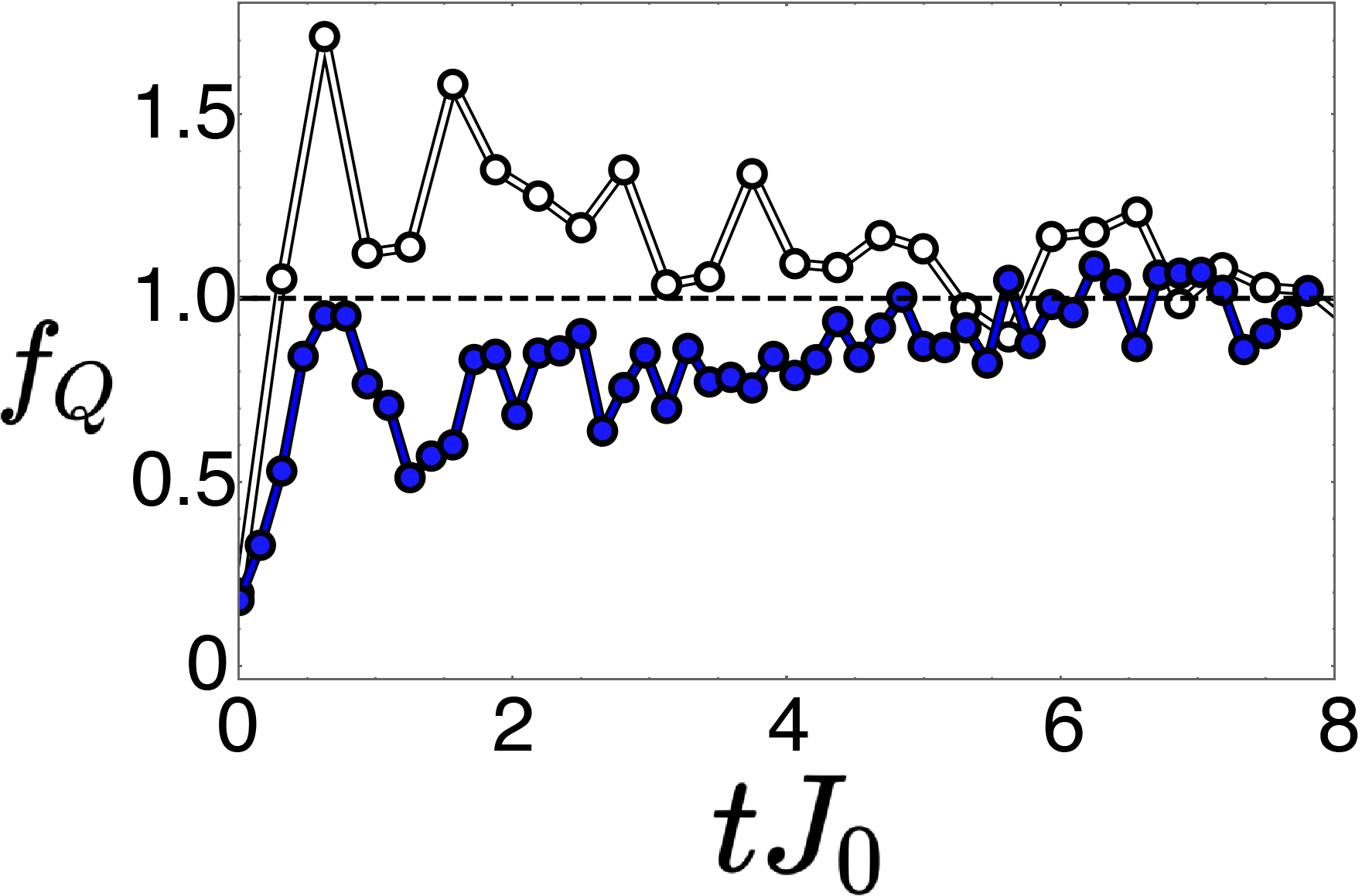}
\caption{Quantum Fisher information. Normalized quantum Fisher information for a N\'eel state $(N=15)$ with $g/J_0=0.24$ (white) and $g/J_0=2.4$ (blue), corresponding to the lowest and highest-gradient data in Fig.~\ref{fig:Fig2}d. Points are experimental observations, with lines as guides to the eye. A value greater than one (dashed line) is an entanglement witness. After the initial fast dynamics up to $tJ_0\approx 1$, the QFI is consistent with saturation for the small gradient, and with slow entanglement growth for the large gradient, with behavior very similar to that previously observed in disordered MBL \cite{Smith2016}.}
\label{fig:FigQFI}
\end{figure}

Quantum Fisher information (QFI) has gained attention as a scalable entanglement witness \cite{Smith2016,Hyllus2012}. For a pure state, it is nothing more than the variance of the witness operator $\mathcal{O}$: $f_Q=4(\langle \mathcal{O}^2\rangle-\langle \mathcal{O}\rangle^2)/N$. For $f_Q>1$, entanglement is guaranteed to be present within the system \cite{Hyllus2012}. As a correlator that carries some information about entanglement, QFI is similar in spirit to measures such as the Quantum Mutual Information \cite{Taylor2019} and the configurational correlator \cite{Lukin2019}.

In the context of the N\'eel state we measure the QFI for a staggered magnetization operator, which reduces to:
\begin{align}
    f_Q&= \nonumber \\
    &\frac{1}{N} \left[ \sum_{jj'}(-1)^{j+j'}\langle \sigma^z_j \sigma^z_{j'} \rangle-( \sum_j (-1)^j \langle \sigma^z_j \rangle )^2 \right] .
\end{align}
The results are shown in Extended Data Fig.~\ref{fig:FigQFI}. We see a significant difference between $f_Q$ with weak and strong field gradients. In a weak gradient, entanglement builds up rapidly before slowly tapering off. In a strong gradient $f_Q$ instead grows slowly, exhibiting similar behavior as expected for entanglement in an MBL phase and in Stark MBL \cite{Schulz2019}.

A few shortcomings limit the value of the QFI. First, it is only easily calculated when assuming a pure state. Second, it can only be interpreted as an entanglement witness when it exceeds one, challenging in a strongly localized phase. Third, unlike the DEER protocol it does not give spatially resolved information. Finally, in a long-range system it can exhibit different scaling than the entanglement entropy \cite{Safavi-Naini2019}. Still, within these limits the QFI behavior is consistent with the expectations for an MBL phase. The QFI dynamics also closely resemble previous observations for disordered MBL \cite{Smith2016}, consistent with expectations that disorder or strong gradients result in similar entanglement spreading.

\section*{Additional DEER Data}

\begin{figure*}[!htb]
\centering
\includegraphics[width=0.75\textwidth]{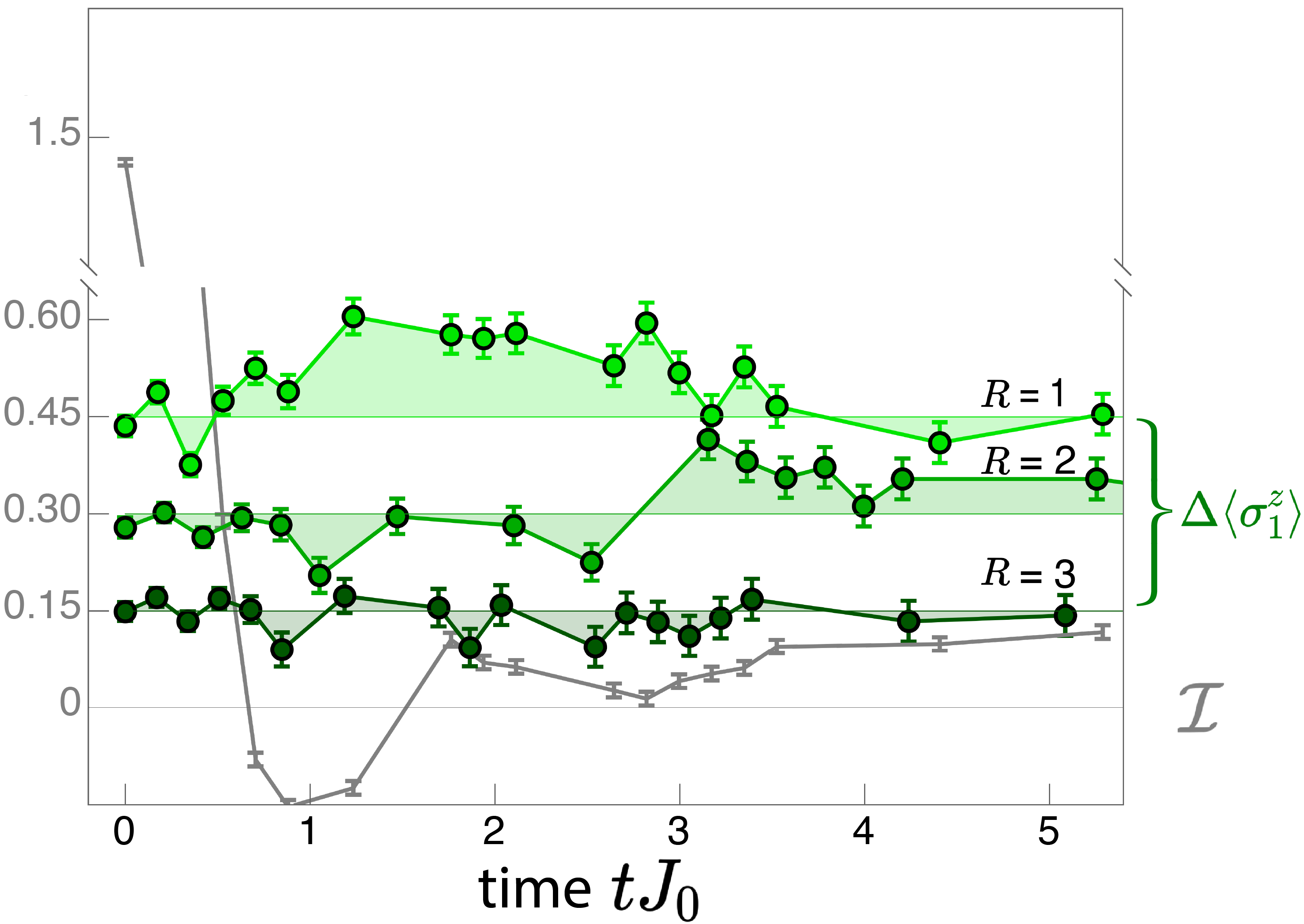}
\caption{DEER Difference signal for $R=$\{1,2,3\} (light to dark), compared with the imbalance $\mathcal{I}(t)$ for the same parameters. Data are offset for clarity but otherwise share the same axes. $\mathcal{I}$ is taken from the same dataset as the $R=1$ spin-echo data, with the probe spin excluded from the imbalance calculation. After $tJ_0\approx 2$, the imbalance is essentially constant at the low but finite steady-state value corresponding to this gradient strength.  However, correlation dynamics are still progressing- in particular, correlations as measured by the difference signal only begin to develop for $R=2$ after this point. This is similar to the disordered MBL state, in which slow entanglement dynamics continue after the locally conserved populations have reached a steady state \cite{Serbyn2013,Huse2014,Lukin2019}.}
\label{fig:FigDEERCompare}
\end{figure*}

Additional data for the DEER protocol difference signal ($\Delta \langle \sigma^z_1 \rangle$) is shown in Extended Data Fig.~\ref{fig:FigDEERCompare}. Looking at the DEER difference signal, we see that correlations develop more slowly as the DEER region $R$ is moved progressively away from the source. For $R=2$, these correlations are only visible after the imbalance dynamics have reached a steady state. This rules out attribution of the correlations to the transient population dynamics, and instead resembles the slow correlation dynamics that occur in a disordered MBL system after populations have reached a steady state \cite{Serbyn2013,Huse2014,Lukin2019}.

\section*{Critical slope in quadratic field}

\begin{figure*}[!htb]
\centering
\includegraphics[width=0.75\textwidth]{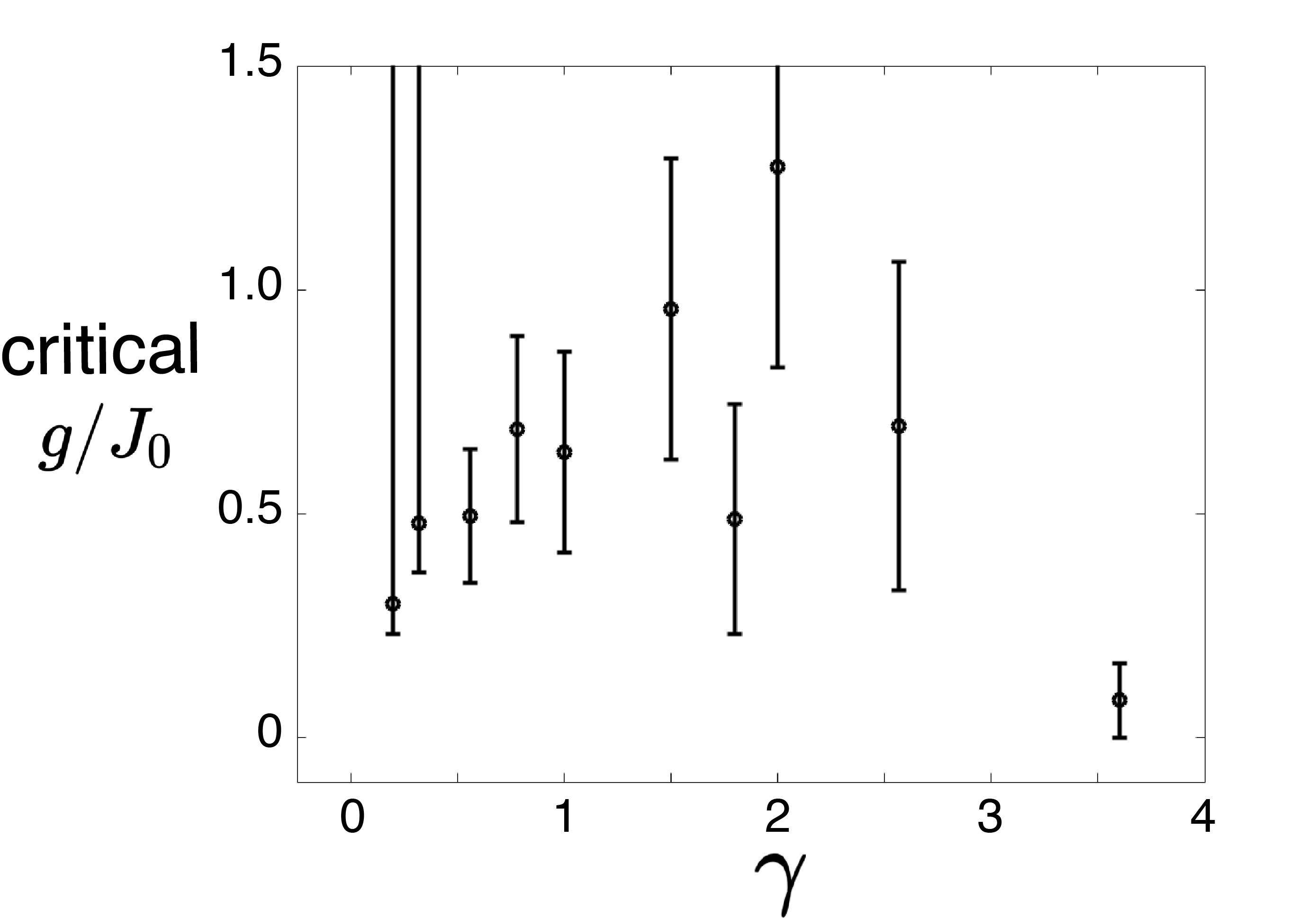}
\caption{Dependence of the critical slope separating thermalizing and non-thermalized regions on the curvature $\gamma$. As the quadratic curvature is varied, the division between thermalizing and nonthermal regions is largely consistent with a critical slope near $g/J_0=0.5$. However, the strongest curvature of $\gamma=3.6$ deviates from this rule. For the lowest two values of $\gamma$ the system was completely delocalized, and thus only the lower bound is meaningful. Error bars (aside from the first two points) denote a variation of $\pm1$ spin location.}
\label{fig:FigQuadCriticalGradient}
\end{figure*}

Extended Data Fig.~\ref{fig:FigQuadCriticalGradient} presents the dependence of the critical value of $g/J_0$ for a quadratic field with different values of the curvature $\gamma.$ The critical value is determined by the innermost pair of spins that are both separated from the center spin by more than their mutual error bars, judged by taking the mean and standard deviation of the average magnetizations for the last five time points.

The data are largely consistent in suggesting a critical gradient value on the order of $g/J_0=0.5$. However, the strongest curvature is notably different, possibly reflecting a breakdown of the local gradient approximation for this case. For curvatures less than this, we conclude that the system seems roughly consistent with a picture of localization that is determined by the local Stark MBL field slope at any given spin.




\end{document}